\documentclass[useAMS,usenatbib]{mnras}
\usepackage{graphicx,natbib,amssymb,amsmath,stmaryrd,makecell}
\usepackage{txfonts}
\usepackage{xcolor}
\usepackage{hyperref}
\usepackage{ulem}

\newcommand{\beq}{\begin{equation}}   %

\newcommand{\eeq}{\end{equation}}   %

\newcommand{\beqa}{\begin{eqnarray}}   %

\newcommand{\eeqa}{\end{eqnarray}}   %

\newcommand{\beal}{\begin{align}}

\newcommand{\enal}{\end{align}}

\newcommand{\bspl}{\begin{split}}

\newcommand{\espl}{\end{split}}

\newcommand{\bsub}{\begin{subequations}}

\newcommand{\esub}{\end{subequations}}

\newcommand{\bmulti}{\begin{multline}}   %

\newcommand{\beqm}{\begin{mathletters}}   %

\newcommand{\eeqm}{\end{mathletters}}   %

\newcommand{\tcr}[1]{{\color{red}{#1}}}
\newcommand{\tcb}[1]{{\color{blue}{#1}}}

\newcommand{\ycobe}{$<\hspace*{-0.1cm}y\hspace*{-0.1cm}>_{\rm COBE}$ }
\newcommand{\yth}{$<\hspace*{-0.1cm}y\hspace*{-0.1cm}>_{\rm T}$ }
\newcommand{\ynt}{$<\hspace*{-0.1cm}y\hspace*{-0.1cm}>_{\rm NT}$ }
\newcommand{\nt}{ntSZ }

\title[non-thermal CMB anisotropies from radio galaxies]{
The non-thermal secondary CMB anisotropies from a cosmic distribution of radio galaxy lobes\\
}

\begin{document}

\author[Acharya et al.]
{Sandeep Kumar Acharya$^1$\thanks{E-mail:sandeep.acharya@manchester.ac.uk},
Subhabrata Majumdar$^2$\thanks{E-mail:subha@tifr.res.in}
and 
Biman B. Nath$^{3}$\thanks{biman@rri.res.in}
\\
$^1$Jodrell Bank Centre for Astrophysics, School of Physics and Astronomy, The University of Manchester, Manchester M13 9PL, U.K.
\\
$^{2}$Department of Theoretical Physics, Tata Institute of 
Fundamental Research, Mumbai 400005, India\\
$^{3}$Raman Research Institute, Sadashiva Nagar, Bangalore 560080, India
}

\date{\vspace{-0mm}{Accepted 2021 --. Received 2021 --}}

\maketitle

\begin{abstract}
Current and upcoming high angular resolution and multi-frequency experiments are well poised to explore the rich landscape of secondary CMB anisotropies. In this context, we compute for the first time, the power spectrum of CMB fluctuations from  a cosmological distribution of evolving lobes of giant radio galaxies.  We, also, explicitly take into account the non-thermal electron distribution,  which has important implications for the inference of the CMB angular power spectrum. We calculate the mean global non-thermal y-distortion, \ynt.  For observationally reasonable distribution of the jet luminosities in the range of $10^{45}-10^{47}$ ergs$^{-1}$, we find \ynt to be less than $10^{-5}$, and hence not violating the COBE limit as previously claimed. Using the unique spectral dependence of the non-thermal SZ, we show that a detection of \ynt can be within reach at the level of  $\gtrsim 5\sigma$ from a future PIXIE-like experiment provided we understand the foregrounds precisely. 
The total non-thermal SZ power spectrum, $C^{NT}_\ell$, from the radio lobes peaks at $\ell \sim 3000$ with an amplitude $\sim 1\%$ of thermal SZ power spectrum from galaxy clusters. A a detection of the $C^{NT}_\ell$, with a PIXIE-like sensitivity experiment, can lead to $\sim 5\sigma$ constraint on the mass dependence of the jet luminosity with the constraint becoming, at least, ten times better for the proposed more ambitious  CMB-HD survey. This will, further, lead to the tightest constrain on the central black hole mass -to- host halo mass scaling relations. 
\\

\end{abstract}


\section{Introduction}
The precise measurement of primary cosmic microwave background (CMB) anisotropy by Planck \citep{Pl2018} has established the standard model of cosmology. Current and proposed CMB experiments, in the post Planck era, have three main goals: (i) measuring the B-mode primordial CMB polarization \citep{KKS1997,HW1997,BICEP2021}, (ii) measuring pre-recombination spectral distortions of the CMB \citep{CS2012,C2021}, and (iii) measuring small angular scale temperature fluctuations (below a few arc minutes) known as the 'Secondary CMB anisotropies' \citep{H2000,AMS2008} caused by the interaction of the CMB photons with the matter in the line of sight. After decoupling with the background electrons, the CMB photons freestream to us. Along the way, they get scattered by the electrons inside the cosmic structure which imprint a characteristic spatial anisotropy and spectral feature which depends on the energetics, densities and velocity fields of these electrons. We define secondary CMB anisotropies to be all temperature fluctuations generated after the decoupling. 

The cosmological sources of secondary CMB anisotropies, as defined above following \cite{AMS2008}, includes: (i) the late time Integrated Sachs-Wolfe (late-ISW) effect \citep{SW1967} due to the CMB photons travelling through a time-dependent gravitational potential in a recent dark energy dominated universe. (ii)the Rees-Sciama effect \citep{RS1968} due to the CMB photons traversing a non-linear gravitational potential as largest scale structures get formed, (iii) the gravitational lensing \citep{LC2006} due to the large scale structure which conserves total power in the temperature fluctuations but transfers power from large angular scales to smaller scales, (iv) the thermal Sunyaev-Zeldovich (tSZ) effect \citep{ZS1969} due to scattering of CMB photons by hot ($\sim$ keV) electrons in cosmic structures, and (v) the kinetic SZ effect (kSZ) \citep{SZ1980,MF2002} due to doppler boosting by bulk flow of electrons which has contribution from both pre-reionization and post- reionization universe \citep{MFHZZ2005,NBTPL2013}. Within the umbrella of SZ effect, different sources of distortions can give rise to a zoo of secondary CMB anisotropies, for example, from quasar outflows \citep{NS1999}, galactic winds \citep{MNC2001}, etc. 


 The thermal SZ effect, due to its redshift independent nature and due to analytic prediction of spectral shape \citep{ZS1969} , has been thoroughly studied \citep{Fixsen1996}. 
 The mean y-distortion and the angular power spectrum of the tSZ effect has emerged as a powerful tool for cosmological and astrophysical constraints   \citep{MS1999,KK1999,B1999,S2001,KS2002,HP2013,Pl20161,BKP2018,ACT2021,SPT2021,HC2021,LCH2021}. The spectral shape also carries information on the energetics of electrons inside these structures which can deviate from the fiducial $y$-distortion due to relativistic electrons \citep{IKN1998,DHPS2001} which can have implications for cosmological analysis \citep{RBRC2019}. Finally, the spectral distortion can also be an ideal tool to detect any departure from the assumption of a thermal nature of the the electrons within cosmic structures \citep{EK2000,M2001}. 

One possible source of non-thermal relativistic electrons could be radio galaxy cocoons/lobes where relativistic particles are supplied by a central supermassive black hole via radio jets \citep{FR1969,S1974,BC1989,N1995}. The energetic particles push out the surrounding gas with the radio lobe expanding from hundreds of kpc to Mpc length scales \citep{KDA1997,AMN2021}. The same population of electrons that are responsible for radio emission also leads to X-ray emission. These have been now detected and can be explained as inverse Compton scattering of CMB photons by the relativistic electrons  \citep{CHHBBW2005,EFB2008,FCCBB2009,JABD2007,ITGHNHSM2009}. One then naturally expects a spectral signature in the CMB, within the frequency range 10-1000 GHz, due to scattering by the non-thermal electrons. Such non-thermal SZ signature from radio galaxies has been studied in \cite{C2008,CM2011} by inferring the energy spectrum of non-thermal electrons from X-ray detection. This, however, requires extrapolation of the energy spectral slope at high energies (inferred from Xrays) down to lower energies. In this regard, a detection of any SZ signature can constrain the low energy spectrum independently. Such a detection of non-thermal SZ has been now been claimed by \cite{MDCMSNW2017}. Recently, \cite{AMN2021} used a physical model for the evolution of radio galaxy lobes and applied it to predict the resultant CMB distortions and radio luminosities. They showed that the above detection of non-thermal SZ, along with the observed radio luminosity, can be used to constrain radio galaxy models as proposed by \cite{KDA1997} and \cite{N2010}. For example, \cite{AMN2021} were able to constrain, for the first time, the low energy threshold of non-thermal electron spectrum ($p_{min}=\sqrt{\gamma_{min}^2-1}$, where $\gamma$ is the Lorentz factor) which is of the order $\sim 1-2$.

In this work, we take the next step and consider the implications for a ubiquitous large scale distribution of radio galaxy cocoons/lobes. Previously, \cite{YSS1999} and \cite{M2001} have looked at the mean spectral distortion signal from a distribution of radio galaxies. Whereas \cite{M2001} finds mean $y$-distortion of $\sim 6\times10^{-7}$ for $\Lambda$CDM universe, \cite{YSS1999} found that results violate the Cosmic Background Explorer(COBE)-Far Infrared Absolute Spectrophotometer(FIRAS) \citep{Fixsen1996} bound. However, the range of values for the luminosity of central engine chosen in that paper appears to be too high \citep{HC2020}. \cite{M2001} also estimated the power spectrum of CMB anisotropy by radio galaxy cocoons and found the Poisson power spectrum to peak at $\ell\sim 4000$. None of the previous papers do a careful calculation of non-thermal spectral distortion, and comparison to COBE-FIRAS limits are thus naive. We revisit these calculations with more reasonable, observationally motivated, radio galaxy models  (Sec \ref{sec:RGmodel}) and accurate use of non-thermal spectral distortion.  Using a halo occupation distribution (HOD) model of radio galaxies (Sec. \ref{sec:radio_galaxy_hod}), we compute globally averaged CMB spectral distortion (Sec. \ref{sec:radio_galaxy_meany}).  We also find the 
magnitude of the mean distortion signal turns out to be significantly smaller compared to \cite{YSS1999} for radio galaxies extending up to $\sim$ Mpc scales. 

We also compute the non-thermal secondary power spectrum (Sec. \ref{sec:radio_galaxy_Cl}) due to distortions of the CMB by these radio galaxies and compare them to thermal SZ power spectrum from galaxy clusters.    
The non-thermal SZ power spectrum  turns out to be  $\sim$ few percent of the SZ power spectrum from galaxy clusters. This indicates that along with other contributions to the total CMB power spectrum, say from clustering of dusty star forming galaxy (refer to top-left panel of Figure 2 in \cite{SPT2021}), the non-thermal SZ power spectrum must be considered for unbiased estimation of cosmological parameters. We look into the impact on cosmological parameters constrained from current SZ experiments in greater detail in a follow-up paper.

Finally, we also discuss the non-thermal distortions in the context of future CMB experiments (Sec. \ref{sec:meany_detection}, \ref{sec:fluctuation_detection}) and show that there is a reasonable chance for their detection. We conclude (Sec. \ref{sec:conclusions}) by doing a rough estimate of the constraining power of such a detection on cosmology, radio galaxy HOD and radio galaxy physics. In the rest of the paper, we use parameters listed in Table \ref{tab:tablecosmo} \citep{Pl2018} as the fiducial cosmological model.

\begin{table*}
   \begin{tabular}{l|r} 
    Cosmological parameters & 1$\sigma$ value   \\
    \hline
    $\Omega_m$ & 0.3111$\pm$0.0056  \\
    $\Omega_{\Lambda}$ & 0.6889$\pm$0.0056  \\
    $H_0$ & 67.66$\pm$0.42\\
    $n_s$ & 0.9665$\pm$0.0038    \\
    $10^9A_s$ & 2.105$\pm$0.030 \\
    $\sigma_8$ & 0.8102$\pm$0.006 \\

\end{tabular}
  \caption{A few of measured and derived cosmological parameters which are most relevant for computing $C_\ell$ of radio galaxies. We have used the values of \citep{Pl2018}.   }
    \label{tab:tablecosmo}
\end{table*}



\section{Model for evolving radio galaxy lobes}
\label{sec:RGmodel}

In our model of radio galaxy, non-thermal relativistic electrons are ejected with a jet luminosity ($Q_j$) and jet lifetime ($t_j$)\footnote{For choice of $Q_{\rm J}$ \& $t_{\rm J}$, see discussion in Section \ref{sec:radio_galaxy_meany}.}
The jet remains active for $\approx 10^7-10^8$ yr after which the jet becomes inactive. The jet pushes out the surrounding gas and forms a radio lobe. The density profile of the surrounding gas can be written as, ${\rho(r)=\Lambda r^{-2}}$, with ${\Lambda}=10^{19}$g cm$^{-1}$ \citep{FMO2004}. The non-thermal electron energy distribution \color{black} in the radio lobe \color{black} is given by a power law with, $n(\gamma_{\rm{i}},t_{\rm{i}})=n_0\gamma_{\rm{i}}^{-\alpha_p}d\gamma_{\rm{i}}$, where $\gamma_{\rm{i}}$ is the Lorentz factor of electrons at time of injection $t_{\rm{i}}$, and ${\alpha_p}$ is the spectral index. The maximum $\gamma$ is assumed to be ${\gamma_{\rm{max}}=10^6}$. The evolution of the radio \color{black} galaxy \color{black} can be described by \citep{RB1997},
  \begin{equation}
  {Q_{\rm J}(t)=\frac{1}{\Gamma_c-1}(V_c \overset{.}{p_c}+\Gamma_c p_c\overset{.}{V_c}) \,, \quad
 \frac{dL_{\rm J}}{dt}=\left(\frac{p_c}{\rho}\right)^{1/2}},
 \label{cocoonsize}
 \end{equation}
 where ${Q_{\rm J}(t)}$ is the jet luminosity which is non-zero when jet is on (${t<t_{\rm J}}$, where ${t_{\rm J}}$ is the jet lifetime) and zero for ${t>t_{\rm J}}$, ${\Gamma_c=\frac{4}{3}}$, ${V_c}$ is the volume of the \color{black}radio lobe \color{black}, ${p_c}$ is the pressure in the \color{black}radio lobe \color{black}, ${L_{\rm J}}$ is the fiducial size of the \color{black}lobe  \color{black} and ${\rho}$ is the density of surrounding gas. We assume the axial ratio of the cylinder shaped \color{black}radio lobe \color{black} to be $\it{R=2}$, the average observed ratio \citep{LW1984}. The volume of the \color{black}lobe \color{black} is then given by,
 \begin{equation}
 {V_c=\frac{\rm \pi}{4R^2}L_{\rm J}^3} \,.
 \end{equation}
 The magnetic and particle energy densities in the \color{black}lobe \color{black} are given by,
 \begin{equation}
 {U_B(t)=\frac{Ap_c(t)}{(\Gamma_c-1)(1+A)} \,, \quad U_e(t)=\frac{p_c(t)}{(\Gamma_c-1)(1+A)}} \,,
 \end{equation} 
 where ${A=(1+\alpha_p)/4}$ \citep{K1997}.
 
 
  The non-thermal electrons cool via adiabatic expansion, synchrotron emission and by inverse Compton cooling off the CMB. The equation for evolution of electron  Lorentz factor ${\gamma}$ is given by,
 \begin{equation}
{\frac{d\gamma}{dt}=-\frac{1}{3}\frac{1}{V_c}\frac{dV_c}{dt}-\frac{4}{3}\frac{\it{\sigma_{\rm{T}}}}{\it{m_{\rm{e}} \rm{c}}}\gamma^2 (U_B+U_C)}\,,
\label{cooling}
\end{equation}
where $\sigma_{\rm{T}}$ is Thomson cross-section, $m_{\rm{e}}$ is the mass of electron, ${c}$ is the speed of light, $U_B$, $U_C$ are the magnetic energy density and CMB energy density, respectively. Due to the $\gamma^2$ factor in Eq. \ref{cooling}, the most energetic electrons in the jet cool very fast. The lowest energy electrons survive for long and these electrons are responsible for creating distortion in the CMB field which has a characteristic shape depending upon the spectrum of non-thermal electrons \citep{C2008,AMN2021}. In this work, we have assumed $\alpha_p=3$, though, the spectral distortion shapes are not sensitive to small changes to $\alpha_p$.


\section{Modeling the distribution of radio galaxies inside dark matter halos}
\label{sec:radio_galaxy_hod}
To calculate the impact of a large scale spatial distribution of radio galaxies on the secondary CMB anisotropies, we need to first compute the cosmological distribution of radio galaxies in the universe. For this, we use the halo model to associate radio galaxies with parent dark matter halos. The abundance of the parent halos is described by the halo mass function (for a particular overdensity $\Delta$) given by,
\begin{equation}
\frac{dn}{dM_{\Delta}}=f(\sigma,z)\frac{\rho_{m0}}{M_{\Delta}}\frac{dln\sigma^{-1}}{dM_{\Delta}},
\label{eq:massfunction}
\end{equation} 
We follow \cite{TKKAWYGH2008, TRKKWYG2010} expression for $f(\sigma,z)$ given by
   \begin{equation}
   f(\sigma,z)=A\left[\left({\frac{\sigma}{b}}\right)^{-a}+1\right]e^{-(c/\sigma^2)}
   \end{equation}
The parameters with their redshift dependence is tabulated in their paper.
\cite{TKKAWYGH2008} provide mass function for different overdensity w.r.t mean matter density while our mass function is defined at overdensity of 500 w.r.t the critical energy density which amount to 500/0.3$\approx$ 1600 overdensity w.r.t to mean matter density. The The value of $r_{500}$ can then be directly computed as $M_{500}=\frac{4\pi}{3}\rho_c\times 500\times {r^{3}_{500}}$. From here on, we drop the subscript $\Delta$ and simply refer to mass function as $dn/dM$. 
Note, that in Sec \ref{sec:radio_galaxy_Cl}, the universal pressure profile used for computing the thermal SZ from galaxy clusters is given  in terms of $M_{500}$. The mass variance $\sigma(z)$ can be computed from linear matter power spectrum with $\sigma(z)=D(z)^2\sigma_{z=0}$. $D(z)$ is the linear growth factor 
\if{
\begin{equation}
\begin{split}
D(z)=\left(\frac{1}{1.0+z}\right)\times 2.5\times \Omega_m(z)\\ \left(\frac{1}{{\Omega_m(z)}^{4/7}-\Omega_{\Lambda}(z)+(1+\frac{\Omega_m(z)}{2})(1+\frac{1}{70}\Omega_{\Lambda}(z))}\right),
\end{split}
\end{equation}
}\fi
normalized to unity at $z=0$. The mass variance $\sigma$ is a function of smoothing scale $R$ and is given by
\begin{equation}
    \sigma(R)=\frac{1}{2\pi^2}\int dk k^2 P(k)\left[\frac{3j_1(kR)}{kR}\right]^2,
\end{equation}
where $P(k)$ is the matter power spectrum which we obtain using publicly available code CLASS \citep{BLT2011}. 


Although the mass function described above gives the abundance of dark matter halos, one still has to associate radio galaxies to their host halos, i.e find a suitable fraction $F(M)$, to get the mean abundance of radio galaxies. In \cite{YSS1999} and \cite{M2001}, a mass independent $F(M)$ was a taken to be 0.01, i.e $1\%$ of the halos in the mass-range $(10^{12}-10^{14})$ M$_\odot$ are assumed  to host radio galaxies. The value of $F(M) = 0.01$ was arrived through looking at the ratio of the total integrated radio luminosity -to- optical luminosity of galaxies with a magnitude cut of $M_V < -22$ as the upper limit.

In our work, we use the halo occupation distribution (HOD) to link radio galaxies to their host halos \citep{WCSJ2008, SFZSTGS2011}. Below, we briefly describe their main results that we adopt and urge the reader to look up their papers for details.
In \cite{SFZSTGS2011}, the radio galaxy fraction is given by $F(M)=A_{\rm HOD}\left(\frac{M}{10^{13.6}M_\odot}\right)^{\beta}$, with the fiducial value of $A_{\rm HOD}=0.09$ and $\beta=0.86$.  As before, we assume that the radio galaxies reside inside dark matter halos with mass $10^{12}-10^{14}M_\odot$ and use this HOD fit in this mass regime. 
A similar radio galaxy fraction was previously given in \cite{WCSJ2008} where
$F(M)=0.02\left(\frac{M}{4\times 10^{12}M_\odot}\right)^{0.65}$. The halo mass in \cite{SFZSTGS2011} is defined w.r.t $r_{200}$ or the radius within which the density of mass is 200 times the critical density. We convert it to $M_{500}$ using the transformation given in \cite{HK2003} and using the concentration relation of \cite{DSKD2008}.

\begin{figure}
\centering 
\includegraphics[width=\columnwidth]{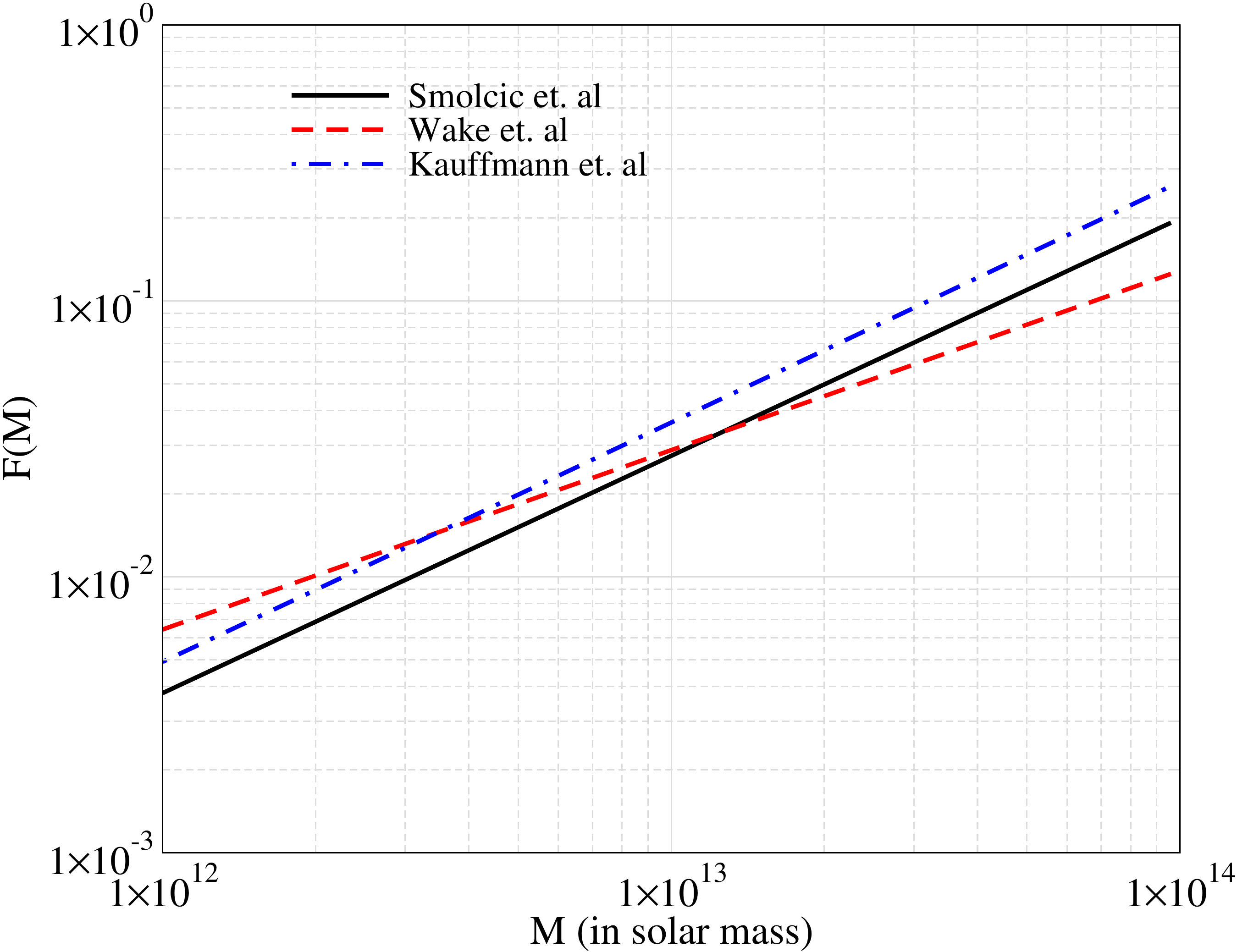}
\caption{The host halo mass dependence of radio galaxy fraction ($F(M)$)}  
\label{fig:radio_hod}
\end{figure}

We also follow another alternate route, based on a modification of the initial approach taken by  \cite{YSS1999}, to estimate the number of radio galaxies inside a halo of mass $M$. The average number of galaxies as a function of dark matter halo has been formulated in semi-analytic galaxy formation models \citep{KCDW1999,BWBBCDFJKL2003}. In these models, the number of galaxies depends upon galaxy type and luminosity. The number of blue galaxies is given by $N_{blue}=0.7$ if $M_{halo}<M_{blue}$ else $N_{blue}=0.7(\frac{M_{halo}}{M_{blue}})^{0.8}$ with $M_{blue}=4\times 10^{12} h^{-1}M_\odot$ \citep{CS2002}. Similarly, the number of red galaxies is given by $N_{red}=(\frac{M_{halo}}{M_{red}})^{0.9}$ with $M_{red}=2.5\times 10^{12} h^{-1}M_\odot$ if $M_{halo}>10^{11} h^{-1}$ solar mass. 
We assume the number of radio galaxies to be $F(M) = 0.01 \times (N_{blue}+N_{red})$ to represent the fraction of number of radio galaxies with respect to sum of total number of red and blue galaxies. We plot the different mass dependent radio galaxy fractions discussed above in Fig. \ref{fig:radio_hod}. They agree with each other within a factor of 2 at the two mass ends, and better within.

\begin{figure}
\centering 
\includegraphics[width=\columnwidth]{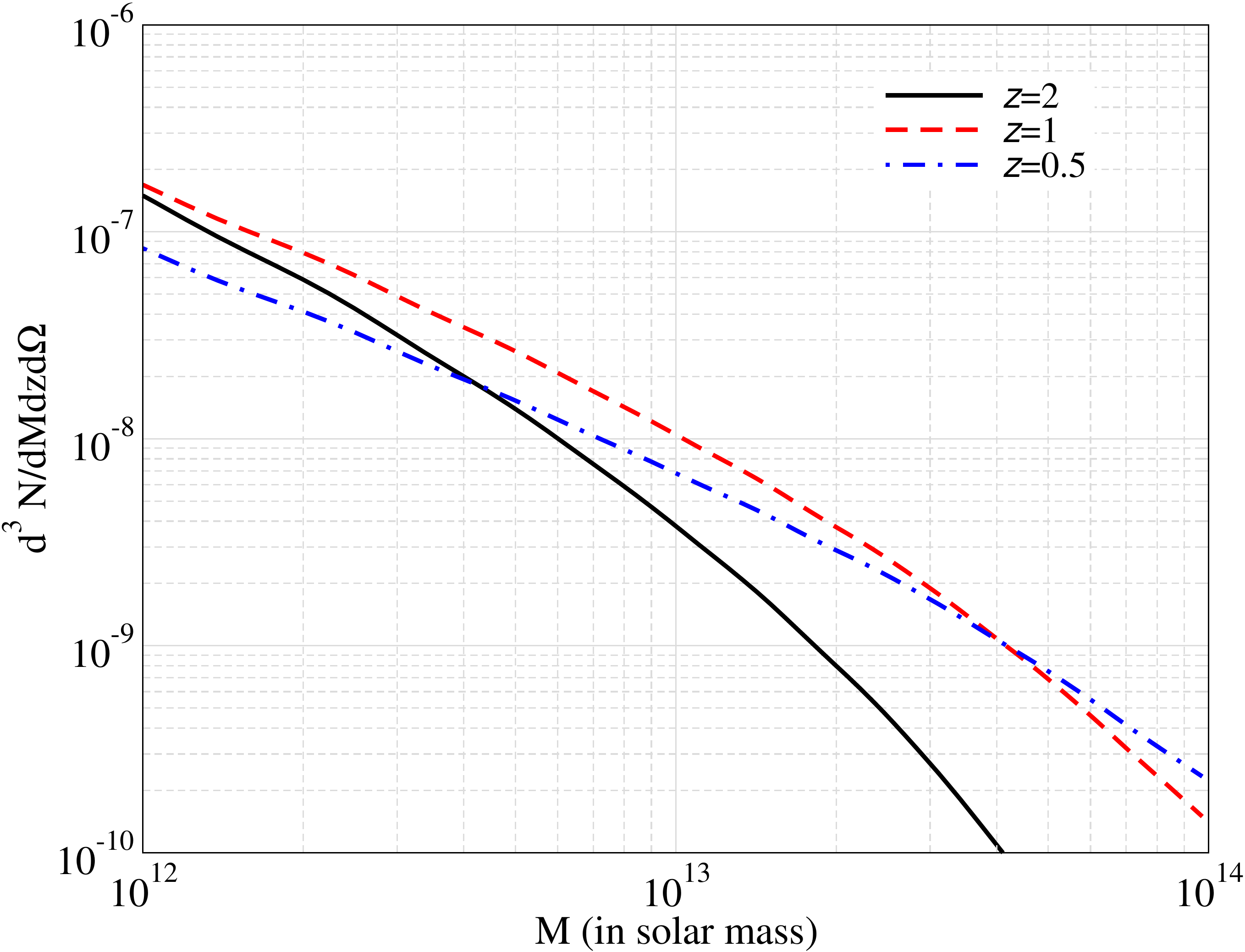}
\caption{Radio galaxy space density,$\frac{d^3N}{dMdzd\Omega}$, as a function of host halo mass for three different redshifts.}
\label{fig:radio_mass_function}
\end{figure}

In Fig. \ref{fig:radio_mass_function}, we plot the $z$-dependent number density of radio galaxies given by
\begin{equation}
    \frac{d^3N}{dMdzd\Omega}=\frac{d^2V}{dzd\Omega}\frac{dn(z)}{dM}F(M),
    \label{eq:radio_mass_function}
\end{equation}
where $\frac{d^2V}{dzd\Omega}=\frac{c\chi^2(z)}{H(z)}$, $\chi(z)$ is the comoving distance to redshift $z$ and $dn/dM$ is the mass function defined in Eq. \ref{eq:massfunction}. We have normalized the matter power spectrum such that $\sigma_8 = 0.81$.  

\section{The global CMB distortion from cosmological distribution of radio galaxies}
\label{sec:radio_galaxy_meany}
\begin{figure}
\centering 
\includegraphics[width=\columnwidth]{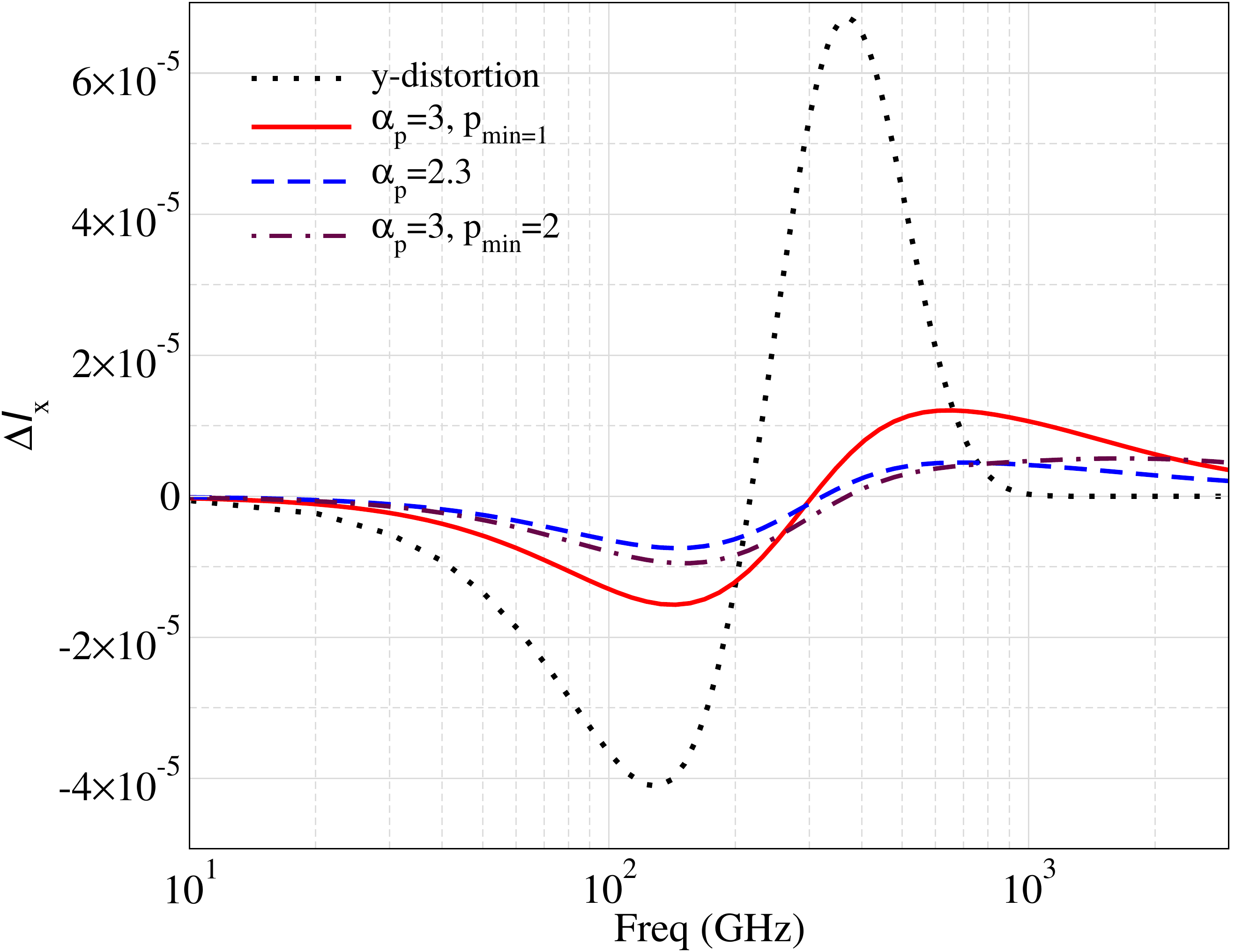}
\caption{Non-dimensional intensity $\Delta I_x(=x^3\Delta n)$ of spectral distortion as a function of frequency $\nu=56.75\times x$ GHz where $x=\frac{\rm{h}\nu}{\rm{k_B}T_{\rm CMB}}$ from thermal and different non-thermal electron population with the same $y$-parameter=$10^{-5}$. Here 
$p_{min}=\sqrt(\gamma_{min}^2-1)$ when $\gamma_{min}$ is the minimum energy and $\alpha_p$ is the spectral index of the non-thermal electron energy distribution is as shown in the plot.}
\label{fig:SZspectrum}
\end{figure}

In this section, we calculate the cumulative global CMB distortion from a cosmological distribution of radio galaxies. To avoid confusion between thermal and non-thermal SZ distortions, we will use $y_{\rm T}$ parameter (or \yth for mean) to denote thermal SZ distortions and  $y_{\rm NT}$ (or \ynt for mean) for non-thermal SZ distortions.
Without doing a full calculation, it is easy to  estimate an upper limit on \ynt. This can be obtained by comparing with the maximum limit on the mean CMB distortions coming from the COBE-FIRAS experiment \citep{Fixsen1996} given by 
\ycobe $\le 1.5\times10^{-5}$ at $2\sigma$. While the mean SZ distortion from galaxy clusters is very close to COBE $y$-distortion limits (with minor relativistic corrections \citep{IKN1998,CL1998,SS1998}), the comparison is straight-forward since the SZ distortion in clusters comes from a thermal population of electrons. In contrast, the non-thermal SZ distortion, in this study, comes from a non-thermal population of electrons and the spectral shapes of CMB distortion from non-thermal -vs- thermal electrons can be vastly different for the same $y$-distortion  \citep{C2008,AMN2021}. More concretely, the distortion in CMB intensity can be written as,
 $\Delta I(x)=\left.y_{\rm{T/NT}} \, g(x)\right|_{\rm{T/NT}}$,  
 where the $y$-parameter (both thermal and non-thermal) depends upon the number density and the spatial distribution of the energetic electrons (which can be same for thermal and non-thermal electrons) whereas the spectral factor $g(x)$ captures the information regarding the energy distribution of electrons. This is shown in Fig. \ref{fig:SZspectrum}, where we compare the spectral distortion shapes for thermal and non-thermal SZ distortions for different non-thermal electron population. Please see also figure 2 in \citep{AMN2021}.

Due to the significant difference in the spectral shapes, the constraint on the non-thermal SZ distortion, when compared with COBE-FIRAS limits, needs to be estimated carefully. The total intensity can be written as \citep{Fixsen1996},
\begin{equation} 
 I(\nu)=B_{\nu}(T_0)+\Delta T\frac{\partial B_{\nu}}{\partial T}+G_0g(\nu)+A\Delta I(\nu),
\end{equation}
where $B_{\nu}(T_0)$ is the Planck spectrum at temperature $T_0$. The second term captures the temperature shift of the CMB i.e. this term is non-zero if the CMB temperature is different from $T_0$. The third term is the galactic spectrum which is given in \cite{Fixsen1996} and the last term is the modelled deviation which in our case will be the non-thermal SZ signal. To proceed, we use the residuals (i.e $I(\nu)-B_{\nu}(T_0)$) given in Table 3 of \cite{Fixsen1996} and fit the linear combination of the temperature shift, our non-thermal spectrum and a galactic spectrum and obtain the best fit values for the corresponding amplitude i.e. $\Delta T, G_0$ and $A$ 
 We obtain a $2\sigma$ upper limit on $y$-parameter for the non-thermal spectrum to be $1.2\times 10^{-4}$. The higher value of the estimated $y$-distortion, compared to thermal $y$-distortion, is expected as the peak intensity of non-thermal spectrum is more than a order of magnitude lower than the thermal spectrum for the same $y$-parameter as can be seen in Fig. \ref{fig:SZspectrum}.

The cosmology and radio galaxy physics dependent \ynt from a cosmological distribution of cocoons is  given by
\begin{equation}
\begin{split}
<y>_{NT}=\int\int\int F(M) \frac{p_c}{m_{\rm{e}}\rm{c^2}}L_J\sigma_{\rm{T}}\frac{dn}{dM}(M,z_{st})\\ \times \frac{L_J^2}{d_A^2}\frac{d^2V}{dzd\Omega}dz dz_{\rm st} dM,
\label{eq:meany}
\end{split}
\end{equation}
where $\frac{dn}{dM}$ is the comoving number density of the dark matter halos which is given by the halo mass function in Eq. \ref{eq:massfunction}, $z_{\rm st}$ is the starting redshift of jet, $d_A$ is the angular diameter distance ($d_A=a\chi$) where $a(z)$ is the scale factor and $\chi(z)$ is the comoving distance, and $\frac{d^2V}{dzd\Omega}$ is the comoving volume element. We consider a cosmological distribution of radio galaxy with $z_{st}$ varying from 0 to 4. 

Motivated by theoretical modeling and observational evidence connecting central blackhole feedback to host halo mass (for example, see \cite{NR2002,IKNM2018}),
we take the jet luminosity to scale linearly with the mass of the dark matter halo such that the luminosity of the radio galaxies inside a dark matter halo with mass $M$ is given by, $Q_J=10^{45}\left(\frac{M}{10^{12}M_{\odot}}\right)^{\alpha_{\rm M}}$ (with $\alpha_{\rm M}=1$) ergs$^{-1}$ unless otherwise mentioned. Therefore, our fiducial radio galaxy luminosity varies between $10^{45}-10^{47}$ ergs$^{-1}$. The highest observed radio galaxy luminosity turns out to be few times $10^{47}$ ergs$^{-1}$ \citep{HC2020}. 

\begin{figure}
\centering 
\includegraphics[width=\columnwidth]{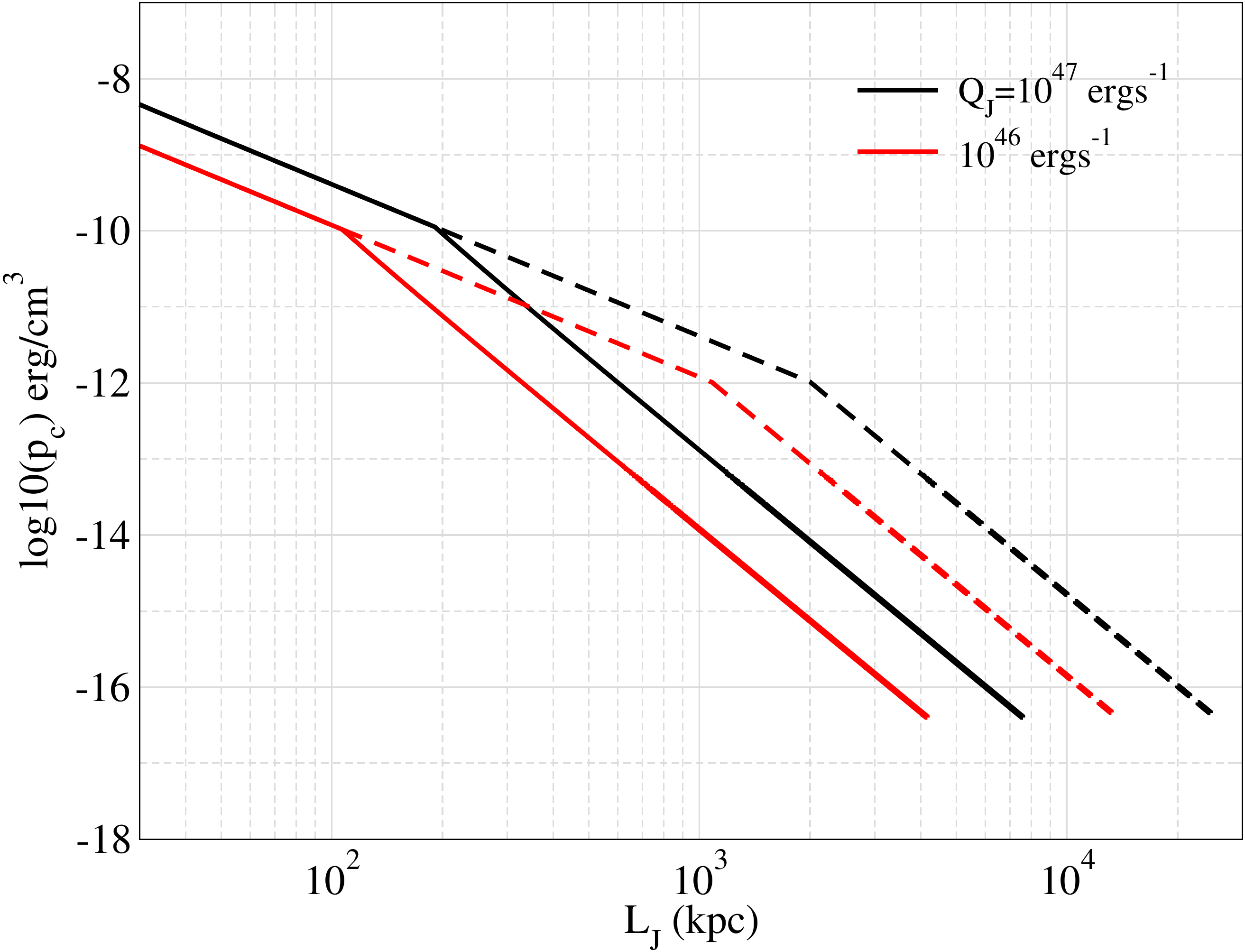}
\caption{Pressure of the non-thermal electrons as a function of the size of radio galaxy lobes. We show the evolution of a radio galaxy with different $Q_J$ and lifetimes $10^7$ yr (solid), $10^8$ yr (dashed) with $z_{st}=1$.}
\label{fig:pressure_vs_Lj}
\end{figure}

Estimates of the duration of  the active phase of jets in radio galaxies, from observations of spectral aging, show that $t_{\rm J}$ lies between $10^7\hbox{--}10^8$ yrs (see, e.g., \cite{KHJC2013}). Also, spectral aging and lobe expansion speed arguments typically estimates  an active timescale of a few $\times 10^7$ yrs (\citep{AL1987,SAAR2008}). Moreover, simulations have also suggested that jets become disrupted after a timescale of order of $10^8$ yr \citep{TD1997,OB2004}.
For the sake of simplicity, we assume that all the radio galaxies have same lifetime and do not have a distribution, and chose our fiducial $t_{\rm J}$ to be $10^8$ yrs. However, we show the dependence of the luminosity and jet lifetime on the pressure as function of the size of the cocoon in Fig \ref{fig:pressure_vs_Lj}.   

\begin{figure}
\centering 
\includegraphics[width=\columnwidth]{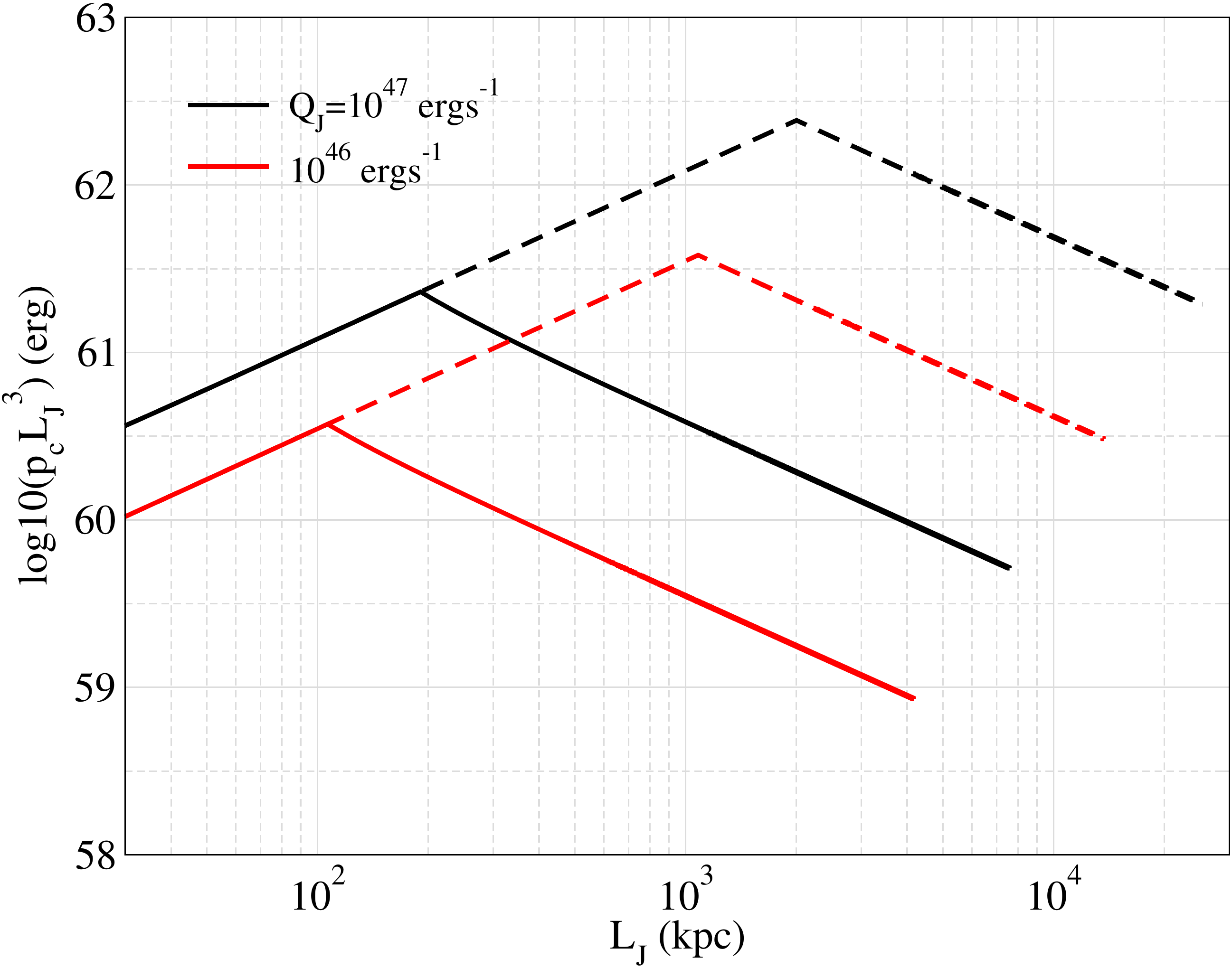}
\caption{The product of pressure and volume of a radio galaxy lobe with same parameters as in Fig.  \ref{fig:pressure_vs_Lj}. }
\label{fig:pcLj}
\end{figure}

Based on a systematic study of radio galaxies using Suzaku, \cite{IK2015} have cataloged radio lobes of varying sizes going up to 5Mpc, which we use as an upper limit on the cocoon size. Note, however, that it is still possible for cocoons to grow to even larger sizes without the hard cutoff.  It is evident from from Fig \ref{fig:pressure_vs_Lj}  that, irrespective of the luminosity or jet lifetime, the cocoon pressure falls by $2-5$ orders of magnitude from its initial value.  
As seen in Fig. \ref{fig:pcLj}, the product of pressure and volume keeps on increasing until the jet stops at which point pressure drops dramatically and the product starts to fall off. Combined with eq. \ref{eq:meany}, it is easy to see that $<y>_{NT}$ receives largest contribution for $\Delta z$ starting with $z_{st}$ and until the jet stops, for one radio galaxy with a given $z_{st}$ or for a cosmological ensemble of radio galaxies. For a given luminosity, the jet expands more for a higher lifetime and the product of pressure and volume is higher (see Fig. \ref{fig:pcLj}). For an ensemble of galaxies, the larger CMB distortion from large cocoons fuelled by larger luminosities are strongly diluted by the exponential fall in the number of higher mass halos. However, some uncertainty due to our incomplete knowledge of jet lifetime remains.

\begin{figure}
\centering 
\includegraphics[width=\columnwidth]{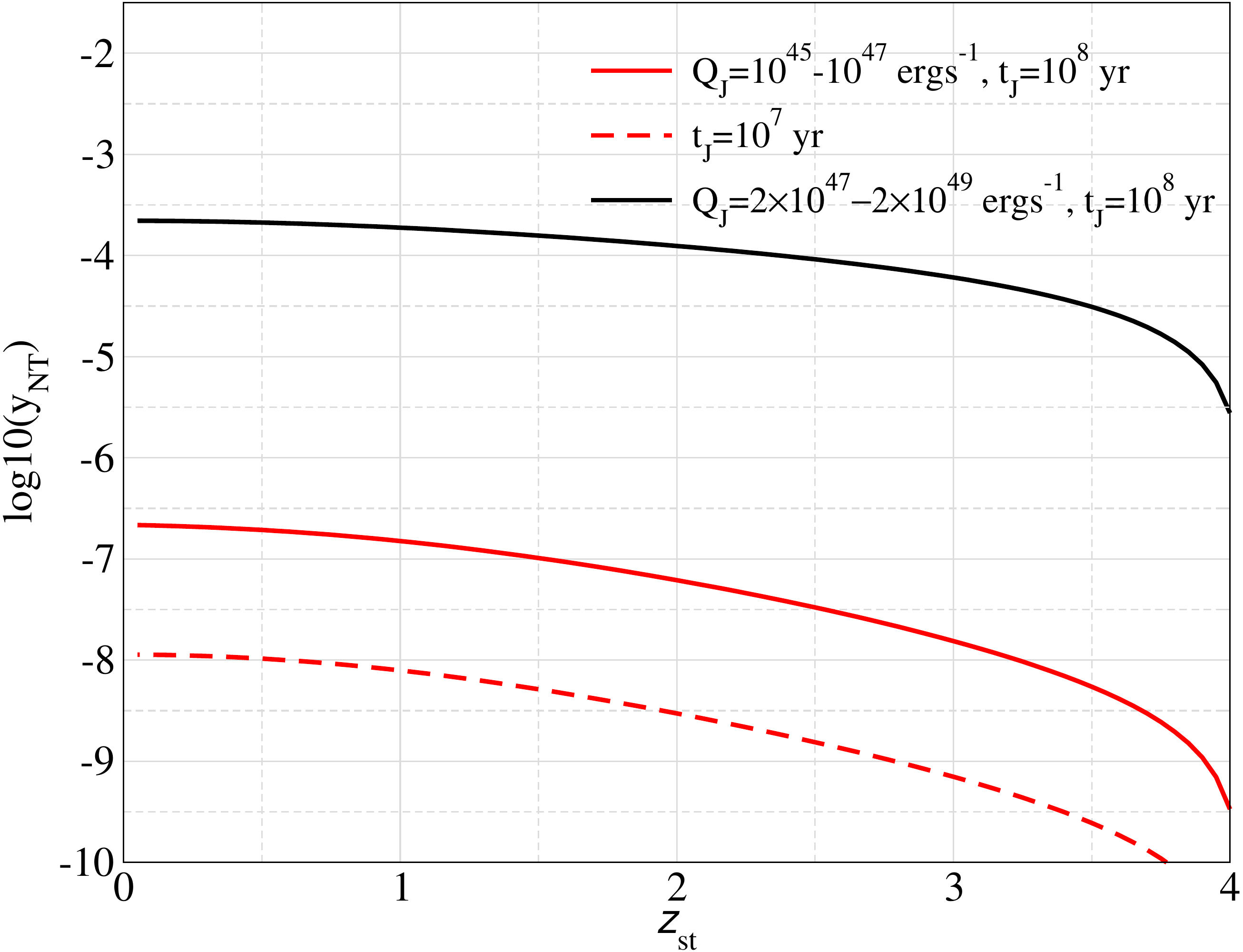}
\caption{The cumulative mean CMB distortion $<y>_{NT}$ from radio galaxies for different jet lifetimes and jet luminosity range.}
\label{fig:y_NT}
\end{figure}

We plot the cumulative $y_{NT}$ from radio galaxies for few different radio galaxy lobes model parameters in Fig. \ref{fig:y_NT}. For comparison with \cite{YSS1999}, we also do a calculation with jet luminosity parameter as chosen by these authors to vary between $2\times 10^{47}-2\times 10^{49}$ergs$^{-1}$. As we increase the jet luminosity and jet lifetime, the pressure increases as expected, and we get larger $y_{NT}$ which can be in conflict with the maximum \ynt estimated above based on COBE-FIRAS limits (for example, see black solid line in Fig. \ref{fig:y_NT}. Using a more physically motivated maximum cocoon size of 5 Mpc, the estimated \ynt is shown in  Fig. \ref{fig:y_NT} which are significantly smaller.  \cite{YSS1999} obtained an estimate for $<y>_{\rm T}$ (and not \ynt) to be greater than $10^{-5}$ for a cosmological distribution of radio galaxies which is in conflict with COBE-FIRAS \citep{Fixsen1996}. With our analytic evolution model of radio galaxies along with recent estimation of the radio galaxy fraction in massive halos, we see that \ynt is several orders of magnitude smaller. Thus, we conclude that there are no constraints on radio galaxy models from COBE-FIRAS \citep{Fixsen1996} data.

\section{Detecting the global CMB distortion}
\label{sec:meany_detection}
\subsubsection{Detecting the non-thermal SZ distortion}

\begin{figure*}
\centering 
\includegraphics[width=1.2\columnwidth]{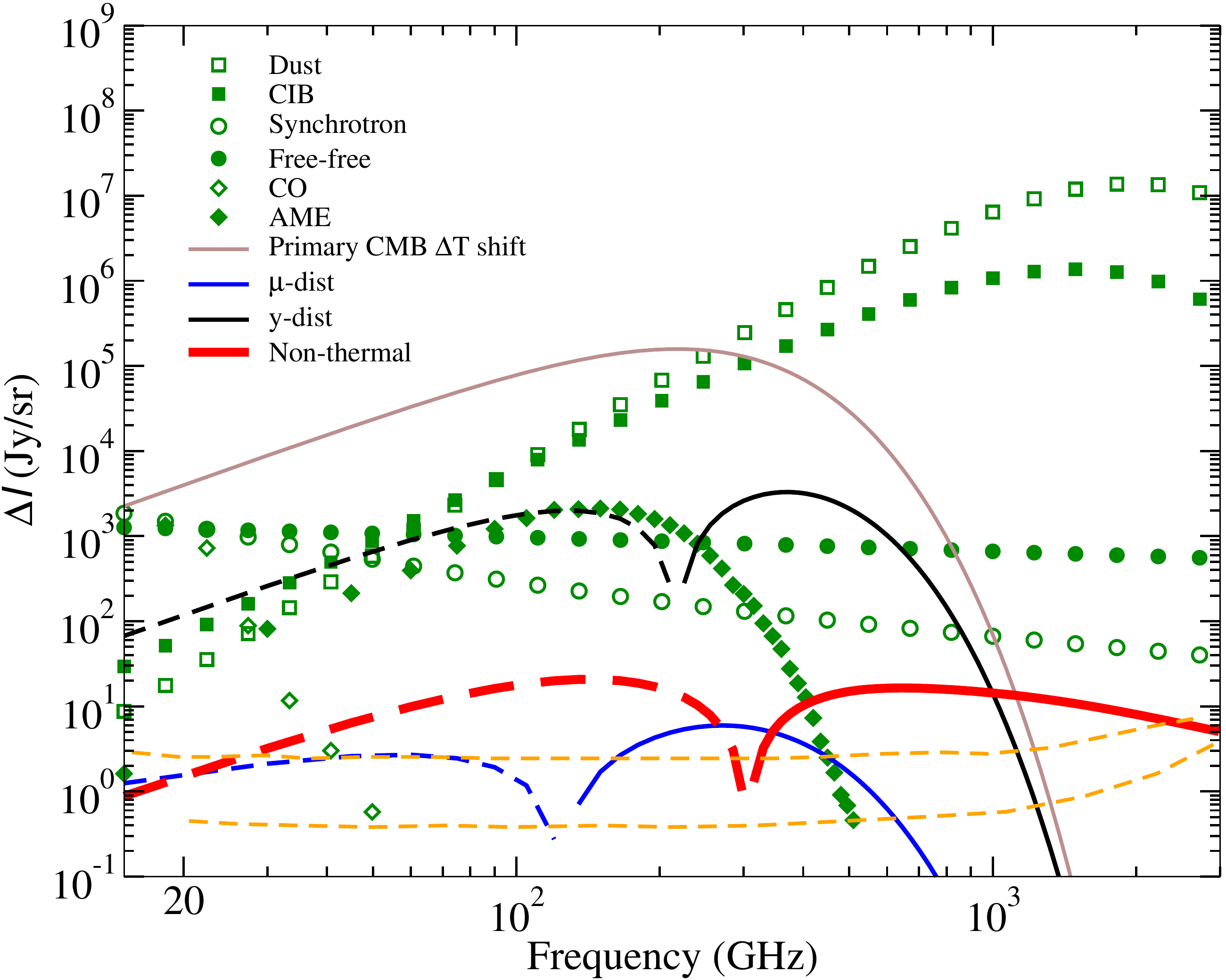}
\caption{The \nt distortion intensity (in Jansky per steradian) and frequency dependence in the context of other cosmological signals and foregrounds. The amplitude of temperature shift, $y$-distortion, $\mu$-distortion and non-thermal SZ are $1.2\times 10^{-4}$, $1.8\times 10^{-6}$, $2\times 10^{-8}$ and $5\times 10^{-8}$ respectively (see text for details). The electron spectral index, $\alpha_p = 3$, used for \nt. The solid and dotted lines for the cosmological signals refer to positive and negative values respectively. We also show PIXIE and super-PIXIE sensitivities with dashed orange lines.
}
\label{fig:foregrounds}
\end{figure*}

\begin{table*}
   \begin{tabular}{l|c|c} 
    Case & Best fit removed & Best fit unremoved   \\
    \hline
    $y+\mu+\Delta T$ & 39$\sigma$  & 39$\sigma$ \\
    $y+\mu+\Delta T$+rSZ & 35$\sigma$  & 35$\sigma$ \\
    Astro foregrounds & 5.38$\sigma$   & 2.8$\sigma$    \\

\end{tabular}
  \caption{Detection forecasts for non-thermal SZ in presence of cosmological and astrophysical foregrounds with PIXIE sensitivity.
  }
    \label{tab:table1}
\end{table*}


After calculating the mean secondary non-thermal SZ distortion ($\rm{ntSZ}$) from a population of radio galaxy cocoons in the last section, we now focus on the prospect of detecting \ynt in future. To proceed, we first plot the predicted cosmological signals and the foregrounds in Fig. \ref{fig:foregrounds}. 
We consider a total of 10 foregrounds, listed in Table 1 of \cite{ACHJ2017} which are divided into two types. The cosmological foregrounds, which we discuss below, are $y$ and $\mu$-distortions, relativistic SZ (rSZ) and temperature shift of CMB black body. The astrophysical foregrounds are dust, cosmic infrared background, synchrotron, free-free, CO and spinning dust (AME).  Since, we are interested at the prospects of a future detection, we focus on the Primordial Inflation Explorer (PIXIE)-like observational program, and also show the instrumental noises for PIXIE \citep{pixie2011} and super-PIXIE. In particular, PIXIE has 400 frequency channels from 15 GHz to 6000 GHz which are equally placed at 15 GHz intervals. However, above 3000 GHz, the experimental noise starts to increase rapidly and, hence, we consider 200 channels from 30 GHz to 3000 GHz in our analysis.

We consider the temperature shift ($\Delta T$) of the CMB spectrum arising from the uncertainty in accurately measure the CMB temperature, $y$-distortion from galaxy clusters and $\mu$-distortion from pre-recombination adiabatic cooling \citep{Sz19701,CS2012} as sources of cosmological signals in standard cosmology. The mean value of \yth from galaxy clusters in the universe is taken to be $\sim 1.8\times 10^{-6}$ \citep{HBCFSS2015}. As the mean temperature of hot gas in these galaxy clusters can be $\sim$1 keV, relativistic correction  \citep{IKN1998,CL1998,SS1998,DHPS2001}  to non-relativistic $y$-distortion becomes appreciable which we also include here. We parameterize this as rSZ corrected $y$-distortion with its amplitude being the same as the $y$-distortion. There can be other sources to $y$-distortion from exotic physics such as dark matter decay and annihilation \citep{SC1984,CS2012,KSC2012,CKS2012}; however, we ignore such non-standard sources in this work. The amplitude of $\mu$-distortion expected from adiabatic cooling of electrons in the standard cosmology is $\sim2\times 10^{-8}$ \citep{C2016}. Another source of CMB distortion in the standard cosmology is from cosmic recombination at $z\sim 1000$ \citep{D1975,SC2009,HRC2020}. The amplitude of the signal is of the order $10^{-9}$ and is expected to be uncorrelated with other cosmological signals due to complicated frequency dependence. We ignore this signal in this work. For the non-thermal SZ signal from radio galaxies, we consider a fiducial \ynt= $5\times 10^{-8}$ with the spectral index of electrons, $\alpha_p$, equal to 3. Having the signal and all the foregrounds in place, we first do a fit of the non-thermal SZ in terms of \yth, $\mu$ and temperature shift of the CMB and throw out the best fit, i.e., we write,
\begin{equation}
    \Delta I_{\rm{ntSZ}}= A_T \Delta I_{\Delta T} + A_y \Delta I_{y_{\rm T}} + A_{\mu}\Delta I_{\mu}+A_{rSZ}I_{rSZ},
    \end{equation}
    where $A_T, A_y$, $A_{\mu}$ and $A_{rSZ}$ are free parameters which can have any values and have nothing to do with the values expected from standard cosmology. By throwing out the best fit from non-thermal SZ, we have a residual non-thermal SZ signal which is independent of the linear combination of the other cosmological signals. We consider this to be the relevant non-thermal SZ signal that needs to be detected and look for its detectability using a Fisher matrix approach. We, further, compare these results with the case when the best fit is not removed.  
    
The expression for Fisher matrix is given by \citep{TTH1997,M2004},
\begin{equation}
    F_{\alpha\beta}=\sum_i \frac{\partial \Delta I}{\partial \Delta \alpha}\frac{1}{\delta I^2(\nu)}\frac{\partial \Delta I}{\partial \Delta \beta},
\label{eq:fisher}
\end{equation}
where, for us, $\alpha,\beta$ are model parameters such as amplitudes of cosmological and astrophysical foregrounds and $\delta I(\nu)$ is the PIXIE instrument sensitivity with the assumption that the noise covariance matrix is diagonal. We are interested in the marginalized constraints on non-thermal SZ signal. The Fisher matrix can be written as, 
$
\begin{bmatrix}
   F & B \\
   B^T & C  \\
\end{bmatrix}
$
where $F$ is the submatrix of parameters that we are interested in and $C$ is the submatrix of the parameters that we want to marginalize over and B is correlation matrix elements with our desired parameter. The marginalized Fisher matrix can then be obtained as  $F^{'}_{\alpha\beta}=F-BC^{-1}B^T$ \citep{M2004}.

Since we are interested in the possible detection of the global non-thermal SZ signal, we marginalize over all other parameters other than $\Delta I_{\rm{ntSZ}}$. For the marginalized Fisher matrix, the signal-to-noise (S/N) is given by $\sqrt{F}$. The resultant S/N are shown in Table \ref{tab:table1} for the scenarios when the best-fit model is subtracted and when it is not. We start with the scenario where there are no astrophysical foregrounds and only cosmological signals ($\delta T$, $y_{\rm th}$ \& $\mu$) are present in additional to non-thermal SZ signal. In this case, we can detect the global CMB distortions by the radio galaxy cocoons at 39$\sigma$ significance. The S/N degrades to 35$\sigma$ when rSZ is added since rSZ has a longer frequency tail. Note, that for low mass clusters ($T_{\rm gas} \lesssim$ 5keV), rSZ is very similar to the $y$-distortion. However, the addition of astrophysical foregrounds, with their large amplitudes, degrade the 
the detection significance considerably leaving with the possibility of $\sim 3\sigma$ detection when best-fit is unremoved, which improves to $5.4\sigma$ if the best-fit is removed. In conclusion, there appears to be very good possibility of detecting the non-thermal SZ signal with a survey with PIXIE sensitivity.

We would like to remind the reader that we have assumed the spectral shape of foregrounds to be known. and have used the SED of \cite{ACHJ2017}. These authors, in turn, have used the results from \cite{Planck2016} themselves. We note that to detect the global ntSZ signal we need very precise understanding of these foregrounds and it is not guaranteed that the SEDs assumed here will suffice to make such a detection.  

While the sky-averaged ntSZ signal is of the order of \ynt$\sim 10^{-7}$, individual radio galaxy can have much higher \ynt. The ntSZ signal from individual sources was studied in our previous paper \citep{AMN2021} where it was noted that these can have \ynt of the order of $10^{-4}-10^{-5}$ (Table 2 in that paper). It was also shown that signals from individual sources can be detected from future surveys such as Simons Observatory \citep{SO2019} and CMB-S4 \citep{CMBS42016}.

\subsubsection{Implications for detection of $\mu$-distortion in the presence of non-thermal SZ}
Spectral distortions in the CMB during the epoch $2\times 10^5\lesssim z\lesssim 2\times 10^6$ \citep{CS2012,KS2012} thermalizes to Bose-Einstein distribution (or $\mu$-distortion) due to efficient scattering between CMB photons and the background electrons which is a photon number conserving process. At higher redshifts, photon non-conserving processes become important which drive the distorted CMB spectrum to a Planck spectrum. In $\Lambda$CDM, $\mu$-distortions are generated due to adiabatic cooling of the baryons and from dissipation of primordial small scale perturbations \citep{HSS1994,CKS2012,KSC2012}. Therefore, detection of $\mu$-distortion probes the expansion history of universe in the radiation dominated era and also probes the initial condition of density fluctuations in the universe.  
Given the importance of the $\mu$-distortion, we check to see whether the likelihood of detection of the $\mu$-distortion gets significantly degraded when we add the ubiquitous non-thermal SZ to be an extra cosmological foreground. 
We start by considering the non-thermal SZ to be the only foreground. In this case, the degradation in the detected $\mu$-signal is negligible when considering 200 bands between 30-3000 GHz for a PIXIE like survey. 
With a lesser coverage of frequency bands, we expect to find more correlation between $\mu$ and non-thermal SZ distortion. However,
even for a survey with fewer ($\sim 20$) bands between 30-300 GHz, the degradation remains marginal when we include all astrophysical foregrounds 

 In fact, with 200 frequency bands, the significance of detection of $\mu$ in presence of cosmological foregrounds is $\sim$ 3$\sigma$ irrespective of whether non-thermal SZ is included or not. 
 We also checked the importance of high frequency channels by including or omitting channels between 2000-3000 GHz. The S/N for detecting $\mu$ in presence of astrophysical foregrounds (including $\rm{ntSZ}$) does improve slightly from 1.25$\sigma$ to 2.3$\sigma$ when more channels are added between 2000 to 3000 GHz.



\section{The nonthermal SZ power spectrum from radio galaxy cocoons}
\label{sec:radio_galaxy_Cl}
In the previous sections, we had looked at the mean non-thermal SZ distortion by radio galaxy lobes. A cosmological distribution of radio galaxies will induce spatial fluctuations in the CMB distortions. 
In this section, we calculate the two point correlation function or power spectrum of \nt  fluctuations from radio galaxies and compare it to the thermal SZ power spectrum from galaxy clusters. As already pointed out, the radio galaxies are assumed to reside within dark matter halos and, hence, their spatial and redshift distribution follows that of their host halos, and can be analytically written in the same lines as for galaxy clusters. We will first write the the expression for galaxy clusters and modify it to suit radio galaxy lobes.

For the galaxy clusters, the thermal SZ power spectrum can be written as the sum of one halo and two halo terms. The expression for one halo term for galaxy clusters is given by \citep{MB2000,M2001,KS2002,HP2013},
\begin{equation}
C_\ell^{1h}=g_{\nu}^2\int dz \frac{d^2V}{dzd\Omega} \int dM \frac{dn(M,z)}{dM}|y_l(M,z)|^2 ,
\label{eq:onehalo}
\end{equation}
where $g_{\nu}$ is the value of $y$-distortion spectral function \citep{ZS1969} at a frequency $\nu$, $y_l(M,z)$ is the $l$-space equivalent of the $y$-distortion each halo, and the other terms have their usual meanings. For $y$-distortion, fractional change in intensity $\frac{\Delta I_{\nu}}{I_{\nu}}=\frac{xe^x}{(e^x-1)^2}\left[x\rm{coth}(x/2)-4\right]$  ($x=\frac{h\nu}{kT_{\rm{cmb}}}$) which when converted to fractional change in temperature is given by, $\frac{\Delta T}{T}=\left[x\rm{coth}(x/2)-4\right]$, which is the factor $g_{\nu}$ in Eq. \ref{eq:onehalo}. The value of $g_{\nu}$ is -2 in the Rayleigh-jeans limit i.e. $x\rightarrow 0$. The corresponding expression for the two-halo term is given by,
 \begin{equation}
   \begin{split}
   C_\ell^{2h}=g_{\nu}^2\int dz \frac{d^2V}{dzd\Omega}\left[\int dM \frac{dn(M,z)}{dM}b(M,z)y_l(M,z)\right]^2 \\ P_{lin}\left(\frac{l}{\chi(z)}\right),
\label{twohalo} 
\end{split}
 \end{equation}
 where $P_{lin}(k)$ is the cosmological matter power spectrum in the linear regime and 
 $b(M,z)$ is the dark matter halos bias \citep{TRKKWYG2010} given by
 $b(M,z)=1+ \left(\frac{\delta^2_c}{\sigma(M,z)^2}-1 \right) / (\delta_c)$
with $\delta_c=1.686$.

For galaxy clusters, the expression for $y_l(M,z)$ is given by
 \begin{equation}
 y_l(M,z)=\frac{\sigma_T}{m_{\rm{e}}c^2}\frac{4\pi r_{500}}{l_{500}^2}\int dx x^2\frac{sin(lx/l_{500})}{lx/l_{500}}P_{\rm{e}}(x),
 \end{equation}
 where $r_{500}$ defines the cluster centric radius within which the mean density of matter is 500 times the critical energy density of the universe, $x=r/r_{500}$ with $r$ being the radial distance from the center of galaxy cluster. We get $r_{500}$ from the fitting formulae of \cite{HK2003}. 
 The characteristic multipole, $l_{500}$, of the pressure profile is given by $l_{500}=a(z)\chi(z)/r_{500}$, where $a(z)$ is the scale factor and $\chi(z)$ is the comoving distance of the cluster at redshift $z$. 
 The NFW halo is characterized by its concentration parameter, $c(M,z)=r_{vir}/r_s$, where $r_s$ is the scale radius and $r_{vir}$ denoted the virial radius of the halo. In this work, we use N-body simulations motivated mass-concentration relation \citep{DSKD2008} given by 
 
\begin{equation} 
 c(M,z)=7.85\left(\frac{M}{2\times 10^{12}M_{\odot}h^{-1}}\right)^{-0.081}(1+z)^{-0.71}
 \end{equation}
 
\begin{figure}
\centering 
\includegraphics[width=\columnwidth]{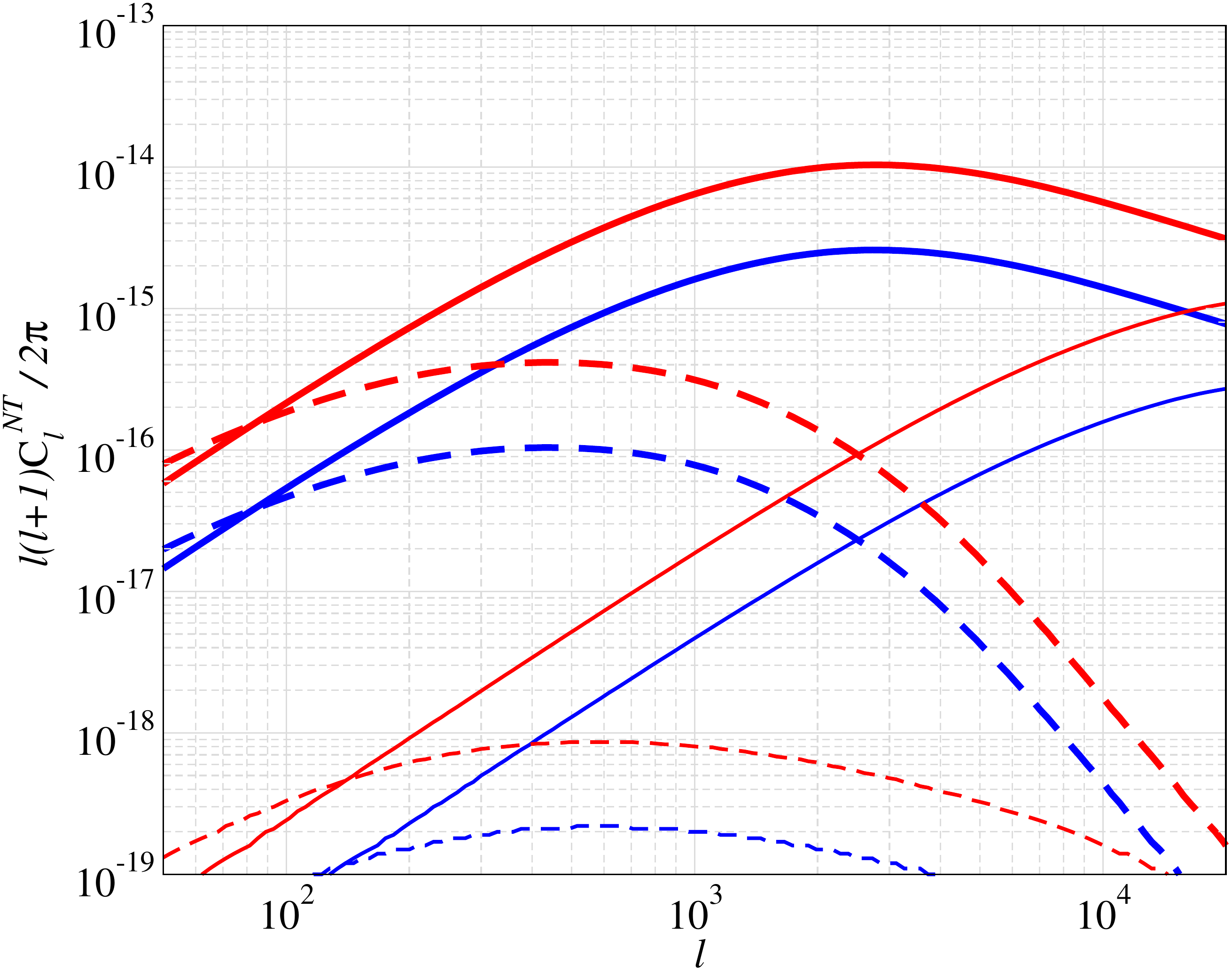}
\caption{The one-halo (solid lines) and two-halo (dashed lines) \nt power spectrum from radio galaxies at 150 GHz. The thicker lines correspond to $t_J=10^8$ yr and the thinner lines are for $t_J=10^7$ yr. For both cases, $\alpha_p=3$  is shown red and  2.3 in blue.}
\label{fig:Cl_radio_one_halo}
\end{figure}

\begin{figure}
\centering 
\includegraphics[width=\columnwidth]{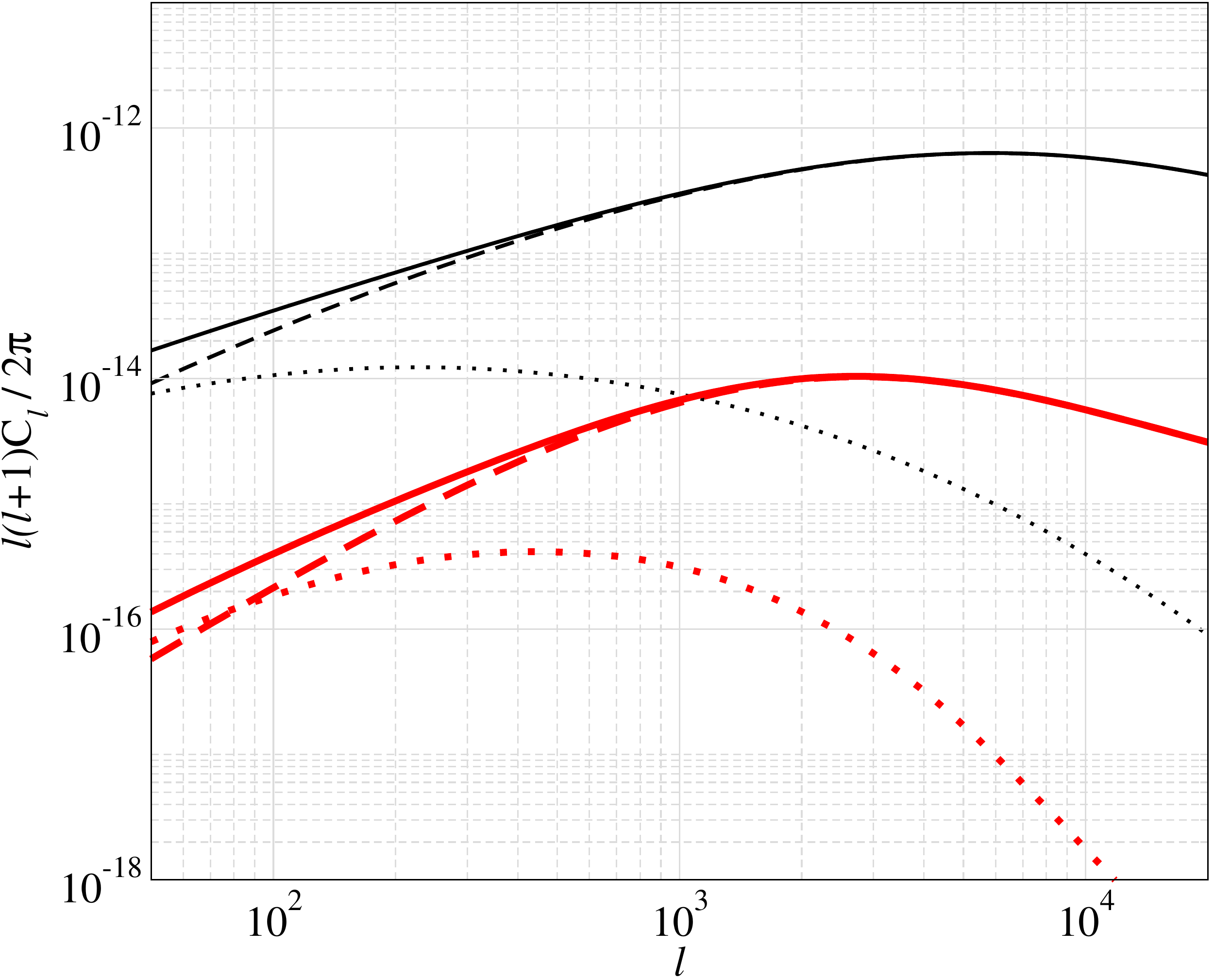}
\caption{Comparison of one-halo (dashed lines), two-halo (dotted lines) and total (solid lines) SZ power spectrum contribution from radio galaxies (thick red lines) and galaxy clusters (thin black lines) at 150 GHz.  Here, $Q_J=10^{45}-10^{47}$ergs$^{-1}$, $t_J=10^8$
yr, $\alpha_p=3$. 
}
\label{fig:Cl_radio_vs_clusters}
\end{figure}

The CMB distortion by galaxy clusters ultimately depends on the pressure, $P_{\rm e}$, of the hot intracluster medium. We use the observationally determined universal pressure profile
 \citep{APPBCP2010,Pl2013} given by
 \begin{equation}
 \begin{split}
   P_{\rm{e}}(x)=1.65\left(\frac{h}{0.7}\right)^2\left(\frac{H(z)}{H_0}\right)^{8/3}\left[\frac{M_{500}}{3\times 10^{14}M_{sun}}\right]^{0.78} \\
   \times P_0(c_{500}x)^{-\gamma}[1+(c_{500}x)^{\alpha}]^{{\gamma-\beta}/\alpha} 
  \end{split}   
   \end{equation}
with the best fit parameters taken from \cite{Pl2013}. Finally, for the cluster SZ power spectrum, we consider dark matter halos in the mass range of $(10^{13}-10^{16})$ M$_\odot$ up to $z\le5$. It is well known that for clusters the 1-halo term dominates at all $\ell$-range, and the total $C_\ell$ peaks at $\ell \sim 3000$. This is seen in Figure \ref{fig:Cl_radio_one_halo} and \ref{fig:Cl_radio_vs_clusters}.

 Next, we turn to the radio galaxy \nt power spectrum ($C_\ell^{NT}$), which has a significantly different spectral shape than thermal $y$-distortion (see Fig. \ref{fig:SZspectrum}). Note again that we assume the radio galaxy lobes to populate dark matter halos in $(10^{12}-10^{14})$ M$_\odot$ mass range  and the jet luminosity is assumed to scale linearly with the mass of the halo. The pressure profile, as discussed in Section \ref{sec:RGmodel}, is assumed to be constant inside the radio galaxy lobe and zero outside. Following the formalism for the galaxy clusters above, we can  compute the one-halo and two-halo terms from Eq. \ref{eq:onehaloradio} and \ref{eq:twohaloradio} below,  
 \begin{equation}
C_\ell^{NT,1h}=h_{\nu}^2 \int dz \frac{d^2V}{dzd\Omega}\int dM F(M)\frac{dn(M,z)}{dM} \int dz_{st} |y_l(M,z,z_{st})|^2,
\label{eq:onehaloradio}
 \end{equation}

 
\begin{equation}
   \begin{split}
   C_\ell^{NT,2h}=h_{\nu}^2 \int dz \frac{d^2V}{dzd\Omega}\left[\int dM F(M)\frac{dn(M,z)}{dM}b(M,z) \int dz_{st}y_l(M,z,z_{st})\right]^2 \\ P_{lin}\left(\frac{l}{\chi(z)}\right),
\label{eq:twohaloradio}
\end{split}
\end{equation}
where the halo mass function is computed at $z_{st}$ and $h_\nu$ is the spectral shape of non-thermal SZ distortion and is different from $g_\nu$ for clusters. 
For $M=10^{12} M_{\odot}$,  $Q_J=10^{46}$ ergs$^{-1}$ and $t_J=10^7$ yr, a jet starting at $z=1$ expands to 1 Mpc by $z=0.9$. For $t_J=10^8$ yr, it reaches the same scale by $z=0.98$. For higher luminosity $Q_J=10^{47}$ ergs$^{-1}$ and $t_J\gtrsim 10^7$ yr, it takes $\Delta z<0.01$ to reach 1 Mpc.  The characteristic size, chosen to be $\sim$ Mpc roughly decides the shape of the power spectrum.

In Fig. \ref{fig:Cl_radio_one_halo}, we compare the one-halo contribution to the SZ power spectrum from radio galaxies to that from clusters at 150 GHz. At this frequency, $g_{\nu}$ of  thermal $y$-distortion$=-0.95$; however, the non-thermal spectrum $h_\nu$ depends on the electron spectrum. We use  $h_{\nu}$ from the non-thermal electron distribution for two different spectral index, $\alpha_{\rm p}$ = (2.3, 3) as shown in the figure.
As seen in the figure,
for luminosities in the range $10^{45}-10^{47}$ erg$s^{-1}$, the 1-halo contribution from radio galaxies is at least an order of magnitude smaller than from galaxy clusters. Higher jet lifetime and a steeper electron spectrum gives greater contribution to $C^{NT}_\ell$.
Since smaller jet lifetimes leads to cocoons of smaller sizes, the peak of the power spectrum moves to higher $\ell$ values.  For $t_J=10^8$ yr, the power spectrum peaks at $l=3000$ or $\theta\sim$ 2-3 arc minutes which can be resolved by CMB experiments such as Atacama Cosmology Telescope (ACT)\citep{ACT2020} or South Pole Telescope (SPT)\citep{SPT2021} with angular resolution of 1.5 and 1 arcminute respectively. We also compare the two-halo terms and see similar pattern as in one-halo term except that the location of peaks of $C^{NT}_\ell$ curves are less visibly sensitive to change in parameters. The total power spectrum (i.e the sum of both 1-halo and 2-halo terms) is shown in
Fig. \ref{fig:Cl_radio_vs_clusters}, where we only consider one population of non-thermal electrons with $\alpha_{\rm p} = 3$. It is seen that the total $C^{NT}_\ell \sim$  1\% of the thermal SZ $C_\ell$ from clusters for most of the $\ell$-range. $C^{NT}_\ell$, quite obviously, depends on both cosmology and radio galaxy physics. For example, for $\ell \gtrsim 2000$, the  $C^{NT}_\ell$ $\propto {\sigma_8}^{4-4.5}$. For the most sensitive radio galaxy parameter $\alpha_{\rm M}$, the $C^{NT}_\ell$ $\propto {\alpha_{\rm M}}^{4.5}$ at its peak, but falls to $C^{NT}_\ell$ $\propto {\alpha_{\rm M}}^{3}$ by $\ell \sim 10^{4}$.

Given the emergence of precision cosmology, it is important for future, higher resolution, CMB experiments to take into account the contribution/contamination from radio galaxy cocoons when constraining cosmology and cluster physics using 
cluster SZ $C_\ell$.  However, impact of the radio cocoon contamination to the cluster thermal SZ $C_\ell$ will be subdominant to other sources of potential systematics arising from uncertain cluster physics. For example, energy feedback from central AGNs \citep{SNBL2010,IMNEEM2017}, electron-ion equilibration and bulk flows of cluster gas (both increasingly important for outer radii) \citep{RN2009} can change the gas profile at large radii and hence the thermal SZ signature at few percent level. More important is the systematic introduced by the mass bias, typically represented by the parameter $b$  which is related to the ratio between the thermal SZ mass $M_{SZ}$, evaluated from the pressure profile assuming hydrostatic equilibrium, and the real cluster mass $M_{500}$ such that $(1-b) \equiv \frac{M_{SZ}}{M_{500}}$. One can calibrate mass bias using additional observations, like weak lensing \citep{Pl20161}, internally using self-calibration \citep{MM2004} or marginalize over $b$ assuming it to be unknown \citep{DSGA2022}. The systematic bias in cosmological parameters ($\sigma_8$, $\Omega_M$) can be as high as ($\sim10\%, \sim 4\%$) depending on putting an external prior on $(1-b)$ versus keeping the mass bias free \citep{DSGA2022} (compare last two columns in their Table 1). 
Moreover, accurate knowledge of completeness and mass cutoff of cluster catalogs (for SZ cluster counts) as well as residuals from projecting out foreground dust and synchrotron components etc (in extracting thermal SZ $C_\ell$)  can also be the sources of biases for precision cosmology with galaxy clusters.
In the future as envisaged in the era CMB-HD observations, once the bias in cosmological parameters due galaxy cluster physics are controlled to percent level, it would necessary to take into account additional systematics introduced from \nt from radio halos.

Till now, we have mainly considered current surveys like ACT and SPT, and the future survey proposal PIXIE. However, the other future high angular resolution and multi-frequency experiments, funded or proposed, are ready to explore the rich landscape of secondary CMB anisotropies. In particular, the CMB-S4 effort\footnote{https://cmb-s4.org/} and newly proposed CMB-HD\footnote{https://cmb-hd.org/} are extremely promising for exploring the CMB distortions discussed in this paper. Keeping this in mind, we also look at the prospects of exploring the \nt from radio galaxy lobes with a CMB-HD like experiments in the next sections.

 \section{Prospects for multi-frequency detection of \nt fluctuations}
 \label{sec:fluctuation_detection}
In this section, we look at the prospect of detecting the cosmological \nt fluctuation power spectrum using multifrequency subtraction of thermal SZ power spectrum from galaxy clusters and other astrophysical foregrounds using the Internal Linear Combination (ILC) method \citep{TED2000}. The central idea is to find a linear combination of frequency maps that minimize the variance of our desired signal. As an example, we consider the case when we want to extract tSZ signal from multifrequency CMB maps ignoring \nt from radio galaxies for the moment.  We start by writing
\begin{equation}
    \hat{a}^{SZ}_{lm}=\sum_{\nu}\frac{w_ia_{lm}(\nu)}{g_{\nu}},
\end{equation}
 where $a^{SZ}_{lm}$ is our estimator for tSZ signal in Rayleigh-jeans limit (where $g_{\nu}\rightarrow 1$), $w_i$ is the linear weight of each frequency map,  $g_{\nu}$ is the tSZ spectral factor and $a_{lm}(\nu)$ is the spherical harmonic coefficients of total temperature anisotropies at frequency $\nu_i$.  The total signal $a_{lm}$ is given by $a_{lm}=\sum_f a^f_{lm}+a^{SZ}_{lm}$, where $f$ stands for foregrounds which are the other astrophysical and cosmological signal except the signal that we want to extract.  We require that the variance $<\hat{a}^{SZ}\hat{a}^{SZ}>$ is minimized subject to $\sum w_i=1$
 The expression of $w_i$ is then given by,
 \begin{equation}
     w_i=\frac{C^{-1}_{ij}e_j}{e_kC^{-1}_{kl}e_l},
     \label{eq:ILC_weights}
 \end{equation}
 where $C_{ij}=\sum_f \frac{C^f(\nu_i,\nu_j)}{g_{\nu_i}g_{\nu_j}}$ is the correlation matrix of foregrounds (i.e. all signals except tSZ in this example) across frequency $\nu_i$ and $\nu_j$ and $e_j$ is a one-dimensional matrix with all entries as 1. If foregrounds (i.e. $C^f_{ij}$) dominate at a particular frequency channel, then the above procedure simply puts a lower weight to that frequency channel to minimize the contribution from the foreground. 
 
 In addition to the foregrounds, we also need to consider the instrumental noise, which after multifrequency subtraction is given by, $N_l=w_iC_{ij}w_j$. After averaging over '$m$'s for each multipole $l$, and for a sky fraction ($f_{sky}=0.7$), the noise is given by $N_l/\sqrt{f_{sky}(2\ell+1)}$. Similar to the noise, the contamination from each foreground can be obtained by replacing $C_{ij}$ by $C^f_{ij}$ in the above expression. 
 The different foregrounds that we consider are those from free-free emission, synchrotron, dust and radio point sources and IR point sources. We use the expression of $C^f_{ij}$ for these components from \cite{TED2000,HP2013}. We list them in the appendix for completeness. Ultimately, one should do a full CMB sky simulation and component separation at the map level especially when data is available. The calculations presented here are supposed to give a rough idea of the detection prospects.

\begin{figure}
\centering 
\includegraphics[width=\columnwidth]{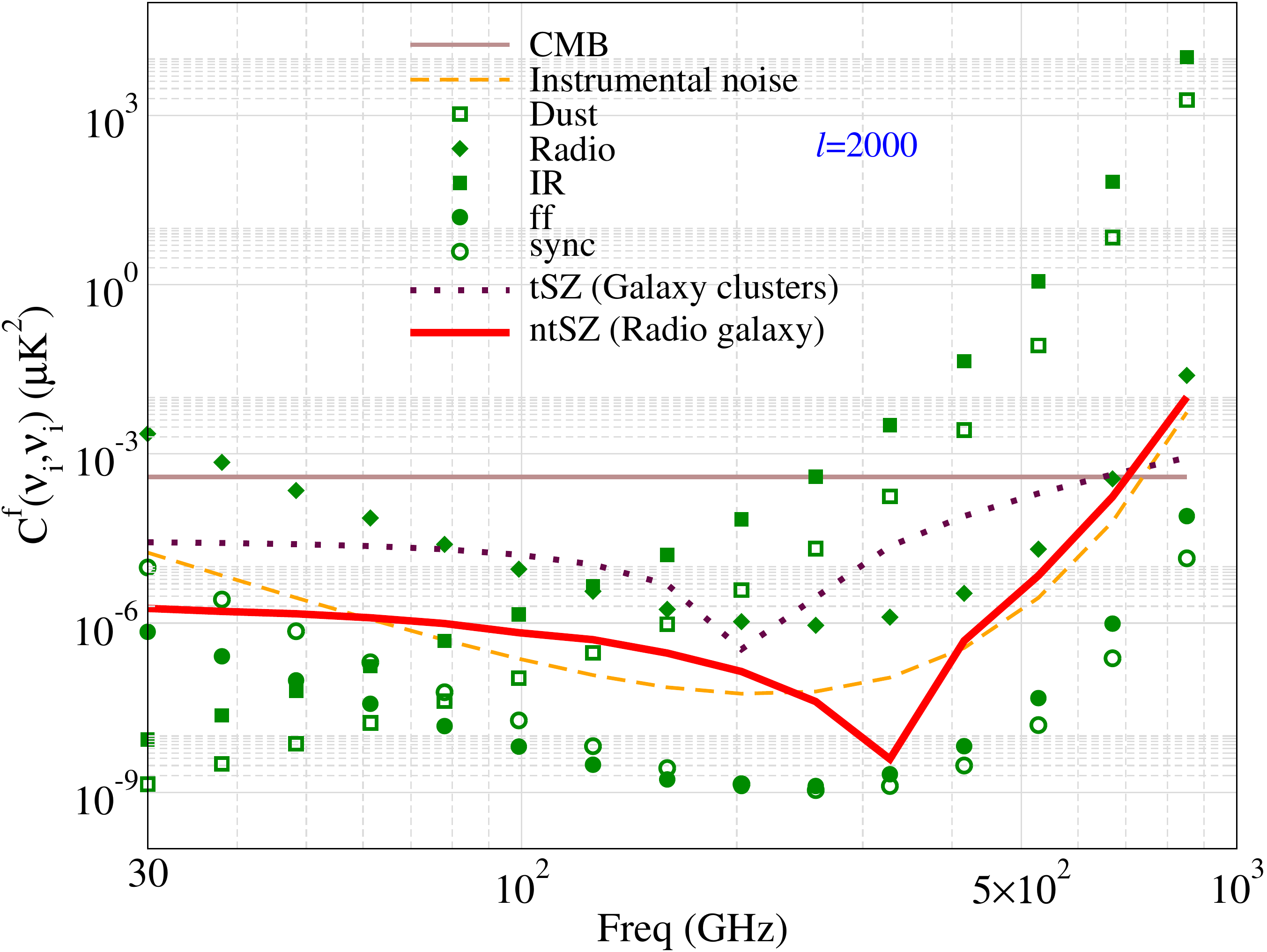}
\caption{Diagonal elements of $C^{f}_l$ as a function of frequency at $l=2000$ for a PIXIE-like experiment. See the text for details. }
\label{fig:noise_pixie_l_2000}
\end{figure}

 Our ultimate goal is to see if \nt signal from radio galaxies, which is roughly 1\% of thermal SZ from galaxy clusters at $l\sim 1000$, can be detected in future. For this purpose, we consider an ambitious "PIXIE-like sensitivity" experiment. Since PIXIE has 400 frequency bands \citep{pixie2011}, the multifrequency subtraction will involve inverting a $400\times 400$ matrix. For exploratory purpose, we consider a  simpler exercise with 15 and 30 frequency bands which are logarithmically spaced between 30 GHz and 850 GHz. We assume that our hypothetical experiment has same noise budget as PIXIE. The sensitivity of PIXIE at each frequency band is $\Delta I=4\times 10^{-24}$ Wm$^{-2}$sr$^{-1}$Hz$^{-1}$ and the resolution is 2.6 degree which translates to $l=\pi/\theta\approx 70$,  beyond which the instrumental noise increase exponentially as can be seen from Eq. \ref{eq:noise} . However, for our hypothetical experiment, we assume that the resolution is of the order of few arc minutes, such that the corresponding $l$ is approximately $5\times 10^3-10^4$.  In Fig. \ref{fig:noise_pixie_l_2000}, we plot the diagonal element of $C^f(\nu_i,\nu_j)$s for each noise component in our experiment at $l=2000$. Over here, the thermal SZ from galaxy clusters are an additional source of foreground (or noise) for the extraction of non-thermal SZ from radio galaxies. .   
 
\begin{figure}
\centering 
\includegraphics[width=\columnwidth]{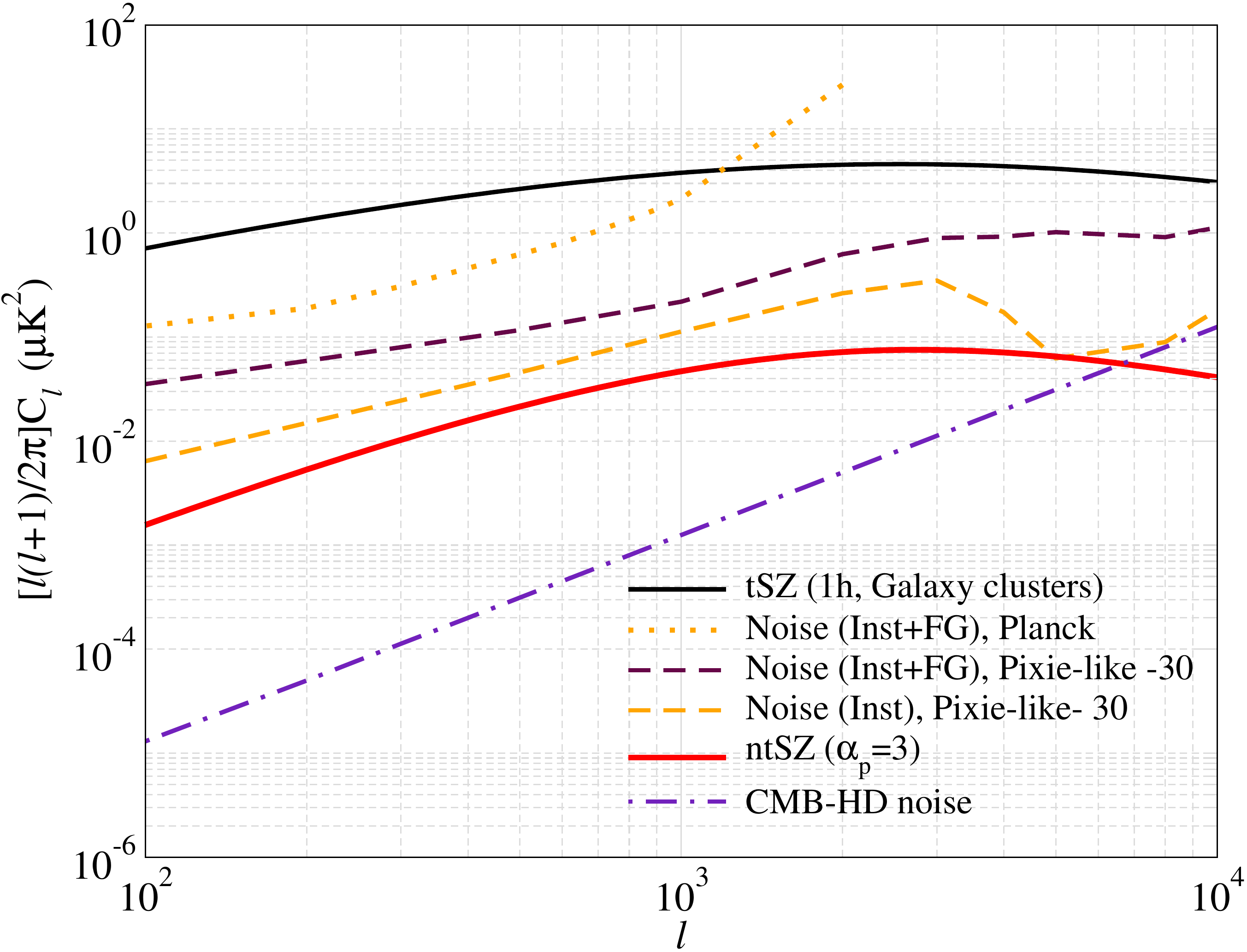}
\caption{Instrumental (Inst) noise (dashed orange), instrumental+foreground (FG) noise (dashed maroon) after applying ILC to extract non-thermal SZ signal from radio galaxies with $Q_J=10^{45}-10^{47}$ergs$^{-1}$, $t_J=10^8$ yr and $\alpha_p=3$ for a PIXIE-like  experiment with 30 frequency channels having $\sim$ arc-min resolution ($\ell\sim 10^4$). We have added Planck noise in dotted orange and tSZ signal from galaxy clusters for reference. (see text for the description). We also show noise for the proposed CMB-HD experiment in dot-dashed violet line.  
}
\label{fig:tsz_radio_extraction}
\end{figure}

 The multifrequency subtraction procedure described here assumes each noise component to be uncorrelated with other. The thermal SZ from galaxy clusters and \nt from radio galaxies may be correlated with each other if they both reside in the same dark matter halos. However, we are interested in \nt from field radio galaxies and ignore such complications in this work. 
The thermal SZ signal is given by an analytical formula and, therefore, the correlation matrix for tSZ across frequency is perfectly correlated just like primary CMB.
Next, we follow the same procedure as done before for extracting tSZ signal with Planck \citep{HP2013} and the result is shown in Fig. \ref{fig:tsz_radio_extraction}. As seen in the figure, we are able to significantly reduce the total noise (shown by dashed maroon) for 30 channels with our hypothetical experiment  compared to Planck (dotted orange line). The reduction of                                                          noise allows us to come within a factor of 2-3 of detection threshold of non-thermal SZ at $l\sim 1000$. At $l$ higher than the resolution of our experiment, instrumental noise dominate. The increase in number of frequency bands, from 15 to 30, does not lead to significant reduction of noise level; however, a much larger number of frequency bands may be able to make a difference. The reduction of overall noise in PIXIE is due to lower instrument noise. Once the noise due to foregrounds starts to dominate, any further reduction of instrumental noise does not make a difference.  This is because the covariance matrix is dominated by few foreground components such as CMB, dust and IR at $l=2000$ in Fig. \ref{fig:noise_pixie_l_2000} which then determines the weights that we obtain after variance minimization in Eq. \ref{eq:ILC_weights}.  These weights may not be optimal for sub-dominant components such as radio point sources and tSZ. Therefore, even if we manage to remove the dominant components with more frequency bands, it does not necessarily make a significant change to cleaning of sub-dominant components. For a more optimized cleaning procedure, one may use constrained ILC (c-ILC) \citep{RDC2011} which puts additional constraints on the unwanted spectral components. Encouraged by the possibility of having S/N within a factor few with our simplistic ILC method, we conclude that the detection of \nt may be within reach with a more optimized analysis for a futuristic PIXIE-like experiment. 

We also do a quick estimate for the proposed CMB-HD survey \citep{CMB_HD2019}, which has a lower noise level than PIXIE (see violet dot-dashed  line in Fig \ref{fig:tsz_radio_extraction}) but with 7 frequency bands between 30-350 GHz. This frequency range may not be enough to distinguish \nt from thermal distortions (see Fig. 11 of \cite{AMN2021} and related discussions) and a few additional high frequency channels ($\gtrsim500$GHz) may be needed to distinguish ntSZ clearly.  Since the CMB-HD proposal is still in its beginning stage, for this work, we assume that there may be additional sufficient frequency coverage to distinguish various distortions.   With CMB-HD, the noise becomes larger than the \nt only above $\ell \sim 7000$, and hence a strong detection of \nt becomes  within reach.

\section{Implications, Discussions and Conclusion}
\label{sec:conclusions}
In this section, we do a fisher matrix estimate on how well we can constrain our radio galaxy physics, parameterized by \{$\alpha_p, t_J, \beta, A_{\rm HOD},\alpha_{\rm M}$\} if $C^{NT}_\ell$ can be extracted from observations. Note, that the $C^{NT}_\ell$s are also dependent on cosmological parameters on which strong priors already exist \citep{Pl2018}. These cosmological priors, along with other conservative priors on the radio galaxy parameters, can be used to break the strong parameter degeneracies.   We list the most relevant cosmological parameters for the determination of $C^{NT}_\ell$ in Table \ref{tab:tablecosmo}. For the present work, we 
fix all the cosmological parameters to their mean values (except $\sigma_8$) since we expect these cosmological parameters to have a negligible effect on $C^{NT}_\ell$ given the precision of their measured values. However, we treat $\sigma_8$ which appears in the exponential in halo mass function as a free parameter and impose Planck prior.

\begin{figure}
\centering 
\includegraphics[width=\columnwidth]{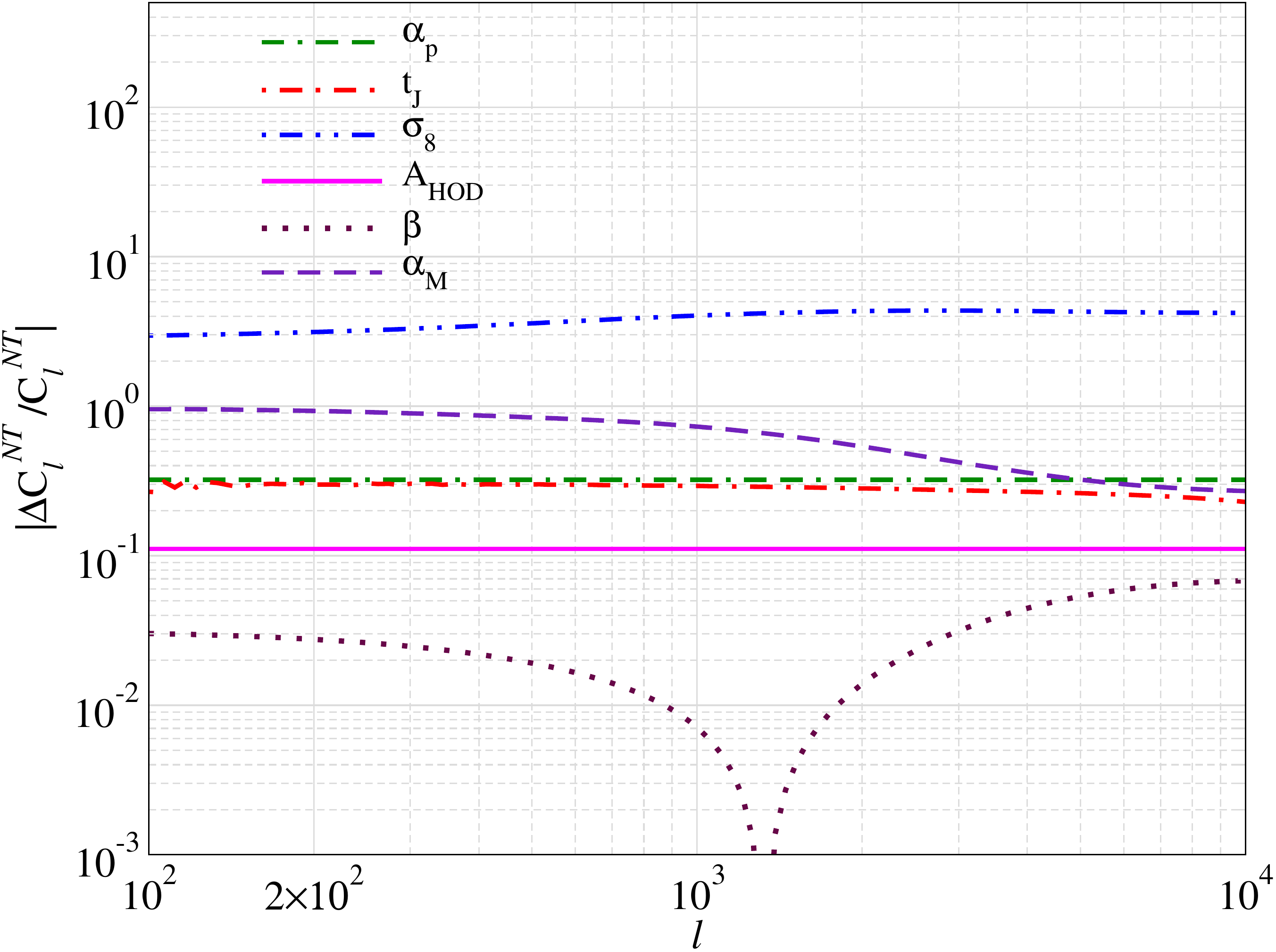}
\caption{Fractional change in the one halo $C^{NT}_\ell$ of radio galaxies due to change in radio galaxy parameters and $\sigma_8$ by 10 percent. The fiducial parameters are  $\alpha_p=3$, $t_J= 10^7$ yr, $\sigma_8=0.81$, $\beta=0.86$, $A_{\rm HOD}=0.09$ and $\alpha_{\rm M}=1$. }
\label{fig:dCl_params}
\end{figure}

\begin{figure}
\centering 
\includegraphics[width=\columnwidth]{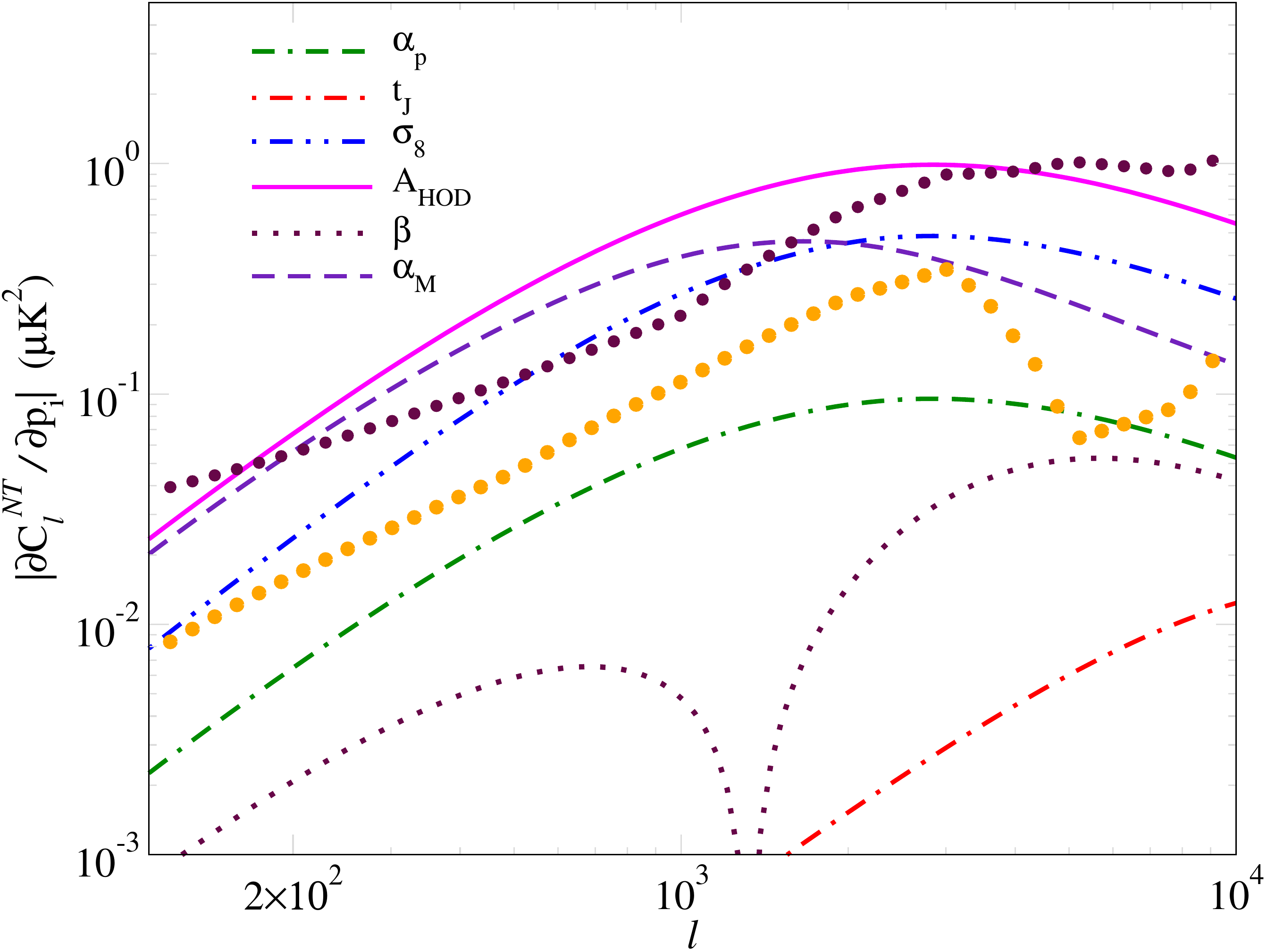}
\caption{$\frac{\partial C^{NT}_\ell}{\partial p_i}$ where $p_i$s are the parameters as denoted in the plot. For $t_J$, the unit of time is per $10^7$ yr. We also show the instrumental noise (orange circles) and total noise(maroon circles) for a PIXIE-like experiment in Fig. \ref{fig:tsz_radio_extraction}.
}
\label{fig:dCl_dp_fisher}
\end{figure}

In Fig. \ref{fig:dCl_params}, we plot the multipole dependence of the fractional changes to $C^{NT}_\ell$, to explore the impact on the $C^{NT}_\ell$ if we change the fiducial radio galaxy parameters and $\sigma_8$.
The fractional changes in $C^{NT}_\ell$  due to change in the fiducial $\alpha_p$, $A_{\rm HOD}$  are independent of $\ell$ and  hence degenerate. This is expected as changing $\alpha_p$ or $A_{\rm HOD}$ simply rescales the $C^{NT}_\ell$. Changing $\sigma_8$ has almost $\ell$ independent behaviour. We have checked that the low-$\ell$ dependence is due to the choice of our high mass cutoff ($M=10^{14} M_\odot$) of dark matter halos hosting radio galaxies.
However, the change to $C^{NT}_\ell$ due to jet lifetime has $\ell$ dependence at high-$\ell$s. For low jet lifetimes, the jet shuts off before the radio galaxy expands to $\sim$ Mpc sizes at which point the pressure profile inside the lobe changes (Fig. \ref{fig:pcLj}) giving rise to the dependency on $l$.  The impact of $\beta$ on fractional change in $C^{NT}_\ell$ also has $\ell$ dependence since a change in $\beta$ changes the relative numbers of larger-to-smaller sized radio galaxy cocoons. 
Changing $\alpha_{\rm M}$ boosts the relative jet luminosity in bigger halos, hence has some scale dependence.  The dip in the curve for $\beta$ is decided by the pivot point of HOD which is between $10^{12}-10^{14} M_{\odot}$.  Finally, in Fig. \ref{fig:dCl_dp_fisher}, we plot $\partial C^{NT}_\ell/ \partial p_i$, i.e the sensitivity of $C^{NT}_\ell$ on the parameters \{$p_i$\} which enters the fisher matrix estimate using Eq. \ref{eq:fisher}.

We  work in the ideal scenario where the foregrounds are removed such that only the instrumental noise (shown in Fig. \ref{fig:tsz_radio_extraction})  remains  and enters the error estimates. For the Fisher calculations, 
we bin the $C^{NT}_\ell$ in 50 logarithmically spaced $\ell$ bins between 100 to 10000. From Fig. \ref{fig:dCl_dp_fisher}, it can be seen that among the radio galaxy parameters, the $C^{NT}_\ell$'s are most sensitive to $A_{\rm HOD},\alpha_{\rm M}$ and the least sensitive to $t_J$.
Given all other parameters fixed except a single parameter of interest, we can expect to determine \{$t_J, \alpha_p, \beta,A_{\rm HOD}, \alpha_{\rm M}$\} at the level of
\{$0.5,\, 4,\,2,\,41,\, 23$\}  $\sigma$s, respectively. For a CMB-HD like survey, the constraints strongly improve to \{$88,583,173,6050,4907 $\} $\sigma$s. However, when all parameters are free to vary, due to degeneracies between the different parameters, the constraints significantly weaken. Some of the degeneracies are broken by putting Planck prior on $\sigma_8$, and further further observationally motivated \citep{SFZSTGS2011}  prior on $A_{\rm HOD}$ and $\beta$ as $\sigma^{\rm Prior}_{A_{\rm HOD}} = 0.03$ and $\sigma^{\rm Prior}_{\beta} = 0.3$.

 \if{
\begin{table*}
   \begin{tabular}{l|r} 
    Cosmological parameters & 1$\sigma$ value   \\
    \hline
    $\Omega_m$ & 0.3111$\pm$0.0056  \\
    $\Omega_{\Lambda}$ & 0.6889$\pm$0.0056  \\
    $H_0$ & 67.66$\pm$0.42\\
    $n_s$ & 0.9665$\pm$0.0038    \\
    $10^9A_s$ & 2.105$\pm$0.030 \\
    $\sigma_8$ & 0.8102$\pm$0.006 \\

\end{tabular}
  \caption{A few of measured and derived cosmological parameters which are most relevant for computing $C^{NT}_\ell$ of radio galaxies. We have used the values of \citep{Pl2018}. }
    \label{tab:table2}
\end{table*}
}\fi

\begin{table*}
   \begin{tabular}{l|c|c|c|c|c|c} 
    Parameters & Fiducial value & Priors &  with $\sigma_8$ prior & with \{$A_{\rm HOD}$, $\beta$\} priors & with  \{$\sigma_8$, $A_{\rm HOD}$, $\beta$\} priors  & with  $\alpha_p$ fixed \& \{$\sigma_8, A_{\rm HOD}$, $\beta$\} priors \\
    \hline
    $ \alpha_{\rm M}$ & 1.0 & - & 0.215 & 0.645 & 0.215 &  0.215\\
    $\alpha_p$ & 3.0 & $ - $ & no constraint &  11.4  &  0.8 &  - \\
    $t_J$ (yrs)   & $10^8$  & - & 1.36$\times 10^8$ & 1.23$\times 10^8$ & 0.72$\times 10^8$ &  $0.68\times 10^8$ \\
     $\beta$ & 0.86 & 0.3  & 3.5 & 0.29 & 0.29&  0.29 \\
    $A_{\rm HOD}$ & 0.09 & 0.03  & no constraint &  0.03  &  0.03&  0.027 \\
    $\sigma_8$ & 0.8102 & 0.006  & 0.006 & 0.0059 & 0.0059 &  0.0059 \\
    \hline
\end{tabular}
  \caption{PIXIE like experiment forecast: 1-$\sigma$ uncertainty on $\sigma_8$ and radio galaxy parameters for different priors. }
    \label{tab:table3}
\end{table*}

\begin{table*}
   \begin{tabular}{l|c|c|c|c|c|c} 
    Parameters & Fiducial value & Priors &  with $\sigma_8$ prior & with \{$A_{\rm HOD}$, $\beta$\} priors & with  \{$\sigma_8$, $A_{\rm HOD}$, $\beta$\} priors  & with  $\alpha_p$ fixed \& \{$\sigma_8, A_{\rm HOD}$, $\beta$\} priors \\
    \hline
    $ \alpha_{\rm M}$ & 1.0 & - & 0.008 & 0.015 & 0.007 &  0.007\\

    $\alpha_p$ & 3.0 & $ - $ & no constraint &  0.57  &  0.37 &  - \\
    $t_J$ (yrs)   & $10^8$  & - & 1.35$\times 10^7$ & 1.23$\times 10^7$ & 1.23$\times 10^7$ &  $1.06\times 10^7$ \\
     $\beta$ & 0.86 & 0.3  & 0.15 & 0.17 & 0.13&  0.13 \\
    $A_{\rm HOD}$ & 0.09 & 0.03  & no constraint &  0.03  &  0.03&  0.016 \\
    $\sigma_8$ & 0.8102 & 0.006  & 0.006 & 0.042 & 0.0059 &  0.0059 \\
    \hline
\end{tabular}
  \caption{CMB-HD like experiment forecast: 1-$\sigma$ uncertainty on $\sigma_8$ and radio galaxy parameters for different priors. }
    \label{tab:table4}
\end{table*}

In Tables \ref{tab:table3} and \ref{tab:table4} , we list the constraints under the influence of different priors for a PIXIE-sensitivity and  a CMB-HD like experiments. As expected, there is negligible improvement in the $\sigma_8$ constraint which is already available from Planck. The radio galaxy parameter constraints follow the trends already evident in Figure \ref{fig:dCl_dp_fisher}, with the mass scaling of the jet luminosity, $\alpha_{\rm M}$, being most tightly constrained. We find that, for PIXIE, $\alpha_{\rm M}$ can be estimated at 4.7$\sigma$ as soon as $\sigma_8$ prior is introduced. CMB-HD, with its higher sensitivity can constraint $\sigma_8$ to $\gtrsim 65\sigma$ even without $\sigma_8$ prior, which further increases to $\gtrsim 125\sigma$ when the $\sigma_8$ prior is added.   It is interesting that $A_{\rm HOD}$ is not constrained with PIXIE even though  the sensitivity of the $C^{NT}_\ell$ to  $A_{\rm HOD}$ lies much above the noise in Figure \ref{fig:dCl_dp_fisher}. This can be understood from Fig \ref{fig:dCl_params} - change in the parameters that results in a near constant $\frac{\Delta C^{NT}_\ell}{C^{NT}_\ell}$ are strongly correlated,
and hence $A_{\rm HOD}$ has strong degeneracies with both $t_{\rm J}$ \& $\alpha_p$. Moreover,
 $A_{\rm HOD}$ \& $\beta$ has direct correlation built in from the definition of $F(M)$.  However, for CMB-HD, one can better the constrain on $A_{\rm HOD}$ to $\gtrsim 5\sigma$, a factor of two improvement over the prior on $A_{\rm HOD}$. Note that $\alpha_p$ is expected to have values within 2.3 and 3 \citep{C2008,CM2011}. Imposing a theoretical prior based on equipartition of magnetic and particle energy densities, one can fix $\alpha_p = 3$. Whereas PIXIE can yield $\sigma_{\alpha_p} \sim 0.8$ and hence cannot differentiate between 2.3 and 3, a CMB-HD like experiment can achieve a factor of two better constraint and prefer one fiducial value of $\alpha_p$ over the other. 
 PIXIE does a poor job on constraining $\beta$ beyond its prior; however, CMB-HD can have $\gtrsim 5\sigma$ constraint of $\beta$, similar to $A_{\rm HOD}$. Along with $Q_{\rm J}$, the jet timescale $t_{\rm J}$ is one of the two main parameters affecting radio galaxy physics models.  
 Due to the low signal to noise ratio for PIXIE, a fiducial $t_J = 10^8$ years can be constrained at $\sim$1.5$\sigma$ (PIXIE) at best; however, CMB-HD has the potential to constrain $t_{\rm J} > 8\sigma$. Within their own capabilities, both PIXIE and CMB-HD like experiments have the potential to  improve our knowledge of radio galaxy physics and their relation to host halos at unprecedented levels.

We have already seen above that, in the presence of Planck priors, the constraint on $\sigma_8$ from just using $C^{NT}_\ell$ is very weak. However, this additional source of SZ distortion can still be  source of bias to $\sigma_8$ constraints using thermal SZ $C_\ell$ from galaxy clusters since at a particular frequency, say 150 GHz, the radio galaxy $C^{NT}_\ell$ can be a percent of cluster $C_\ell$ for $\ell > 1000$. We, of course, need to be careful to include the contribution from the spectral factor correctly. At 150GHz, $g_{\nu}$ for tSZ is -0.95 whereas $h_{\nu}$ for \nt is -0.4 (for $\alpha_p=3$). Hence, the relative spectral boost of the tSZ compared to \nt is $0.95^2/0.4^2=5$. Since, the typical dependence of the SZ power spectrum on $\sigma_8$ goes approximately as $C_\ell \propto \sigma_8^{6-8}$, the impact of miscontributing the \nt from radio galaxies to thermal SZ from clusters could bias $\sigma_8$ to percent level. Although this bias appears small, given that current constraints on $\sigma_8$ from both primary CMB \citep{Pl2018} as well ground based SZ observations \citep{SPT20211} have reached sub-percent level accuracy, it is important to investigate the impact of radio galaxy \nt contamination in more detail taking into account the degeneracies between cosmology and the radio galaxy model parameters. 

In this work, we have considered the \nt from radio galaxies and have not focused on any additional distortion due to kinematic SZ (kSZ) from the bulk motion of the radio galaxies. The
kSZ power spectrum depends upon the product of electron number density and velocity. For a radio lobe, as the lobe expands, the motion of electrons coming towards us and going away from us cancel each other out and only the bulk motion of radio galaxy matters. Assuming equipartition of energy inside the radio lobes, the number density of electrons inside the lobe is proportional to $p_c/m_{\rm{e}}\rm{c^2}$ \citep{KDA1997}, where $p_c$ is the pressure inside the lobe. Therefore, the expression for kSZ power spectrum for radio lobes is equivalent to tSZ power spectrum modulated by $v_{rms}^2$. Since, 
$\frac{v_{rms}}{\rm c} \sim 10^{-3}-10^{-4}$ over the redshift range under consideration, the kSZ power spectrum is 6-8 orders of magnitude smaller than the \nt power spectrum and can be ignored.

\if{ 
\begin{figure}
\centering 
\includegraphics[width=\columnwidth]{./eps/fisher_alpha_sigma8.pdf}
\caption{Fisher matrix constraints on $\alpha_p$ vs $\sigma_8$ with fiducial values -3 and 0.81 respectively, and $t_J, \beta$ fixed at $10^8$ yr and 0 respectively.}
\label{fig:fisher_alpha_sigma8}
\end{figure}
 
\begin{figure}
\centering 
\includegraphics[width=\columnwidth]{./eps/fisher_alpha_tJ.pdf}
\caption{Fisher matrix constraints on $\alpha_p$ vs $t_J$ with fiducial values -3 and $2\times 10^7$ yr respectively, $\beta$ fixed at 0.}
\label{fig:fisher_alpha_tJ}
\end{figure}
 
\begin{figure}
\centering 
\includegraphics[width=\columnwidth]{./eps/fisher_alpha_beta.pdf}
\caption{Fisher matrix constraints on $\alpha_p$ vs $\beta$ with fiducial values -3 and 0.86 respectively. Both $t_J$ and $\sigma_8$ are varied.}
\label{fig:fisher_alpha_beta}
\end{figure}
 }\fi

This paper deals with the \nt caused by the non-thermal plasma in the lobes of radio galaxies fuelled by jets from the central black hole. However, a small part of the energy of the jets also goes into heating the surrounding medium which can lead to the thermal SZ distortions. 
The amount of thermal distortion of the CMB from radio galaxy lobes, to the best of our knowledge, is still described in \cite{EK2000, M2001}. This 
thermal SZ distortion arises from the shocked gas at the cocoon periphery separated from the non-thermal plasma by a contact discontinuity; however, a major uncertainty is the thickness of this thermal envelope. Following \cite{M2001}, the thermal heating can be estimated by calculating the work done by the evolving cocoon on the surrounding medium, similar to what has been presented in section \ref{sec:RGmodel}. A part of this thermal energy is lost by inverse Compton scattering and is  responsible for the thermal SZ distortion.  The efficiency of the inverse Compton loss $f_{IC}$  can be estimated by comparing different cooling timescales, 
$f_{iC} = \frac{1}{t_{iC}} / \left( \frac{1}{t_{ad}}+\frac{1}{t_{ff}}+\frac{1}{t_{iC}}\right)$ where ${t_{ad}}, {t_{ff}} \,\&\,{t_{iC}}$ are the cooling timescales for adiabatic expansion, Bremsstrahlung and inverse Compton, respectively. The cumulative energy supplied to the CMB  and the resultant thermal y-distortion are given by
\begin{eqnarray}
\Delta U_0 &=& \int\int\int F(M) \frac{dn}{dM} \frac{d^2V}{dzd\Omega} \left[f_{iC}(z) \frac{p \frac{dV}{dz}}{(1+z)^4}\right] dz dz_{\rm st} dM \nonumber \\ 
<y>_{th} &=& \frac{1}{4} \frac{\Delta U_0}{aT_{CMB}^4},
\label{eq:szthermal}
\end{eqnarray}
where $a$ is Stephen constant and $T_{CMB}$ is the present day temperature of the CMB.
The resultant thermal SZ distortion is of the order of few times $10^{-8}$ or smaller \footnote{This is much lower than the relatively simpler back-of-the-envelope estimate presented in \cite{EK2000}}. Hence, it is subdominant to the non-thermal SZ from radio cocoons.

 \section*{Acknowledgements}
 The authors would like to thank Jens Chluba and Rishi Khatri for discussions.
SKA is supported by the ERC Consolidator Grant {\it CMBSPEC} (No.~725456).
SM acknowledges support of the Department of Atomic Energy, Government of India, under project no. 12-R\&D-TFR-5.02-0200.

\section{Data availability}
The data underlying in this article are available in this article.


{\small
\vspace{-3mm}
\bibliographystyle{mn2e}
\bibliography{cocoon}
}

 \appendix
 
 \section{Foregrounds and Instrumental Noise}
 The following expression for noise and foregrounds have been used \cite{HP2013}.
 The experimental noise is given by,
 \begin{equation}
 C^N_l(\nu,\nu')=\delta_{\nu,\nu'}\Delta T(\nu)^2e^{l(l+1)\theta(\nu)^2}(8\rm{ln2})\theta(\nu)^2
 \label{eq:noise}
 \end{equation}
 The value of $\Delta T(\nu)$ and $\theta(\nu)$ for Planck experiment can be found in Table 1 of \cite{HP2013}. For the cosmological foregrounds, we parameterize $C^f$ as,
 \begin{equation}
     C^f(\nu_i,\nu_j)=\frac{\Theta^f(\nu_i)\Theta^f(\nu_j)}{\Theta^f(\nu_0)^2}R(\nu_i,\nu_j)C_\ell,
 \end{equation}
 where $\Theta(\nu)$ is the frequency spectrum of a particular foreground and $R(\nu_i,\nu_j)$ is the correlation factor across the frequency whose expression is $R(\nu_i,\nu_j,\alpha)=\rm{exp}\left[-\frac{1}{2}(\rm{log}(\nu_i/\nu_j)/\zeta^f)^2\right]$ with $\zeta^f=(\sqrt{2}\alpha)^{-1}$. The instrumental noise is perfectly uncorrelated across frequency while the CMB is perfectly correlated. Other cosmological foregrounds lie in between.
 
 The frequency distribution is given by,
 \begin{equation}
     \Theta^{CMB}(\nu)=1.
 \end{equation}
 \begin{equation}
     \Theta^{ff}(\nu)=\nu^{-2.15}c(\nu)
 \end{equation}
 \begin{equation}
     \Theta^{Dust}(\nu)=\frac{c(\nu)\tilde{c(\nu)}\nu^{4.7}}{e^{\frac{\nu}{56.8}\frac{2.725}{18}}-1}
 \end{equation}
 \begin{equation}
     \Theta^{sync}(\nu)=\nu^{-2.8}c(\nu)
 \end{equation}
 \begin{equation}
     \Theta^{radio}(\nu)=\nu^{-0.5}\left[\frac{\partial I(\nu,T)}{\partial T}\right]^{-1}_{T_{\rm{CMB}}}
 \end{equation}
 \begin{equation}
     \Theta^{IR}(\nu)=\nu^{2.1}I(\nu,9.7 \rm{K})\left[\frac{\partial I(\nu,T)}{\partial T}\right]^{-1}_{T_{\rm{CMB}}}
 \end{equation}
 with $c(\nu)=\left[\frac{2\rm{sinh}(x/2)}{x}\right]^2$, where $x=\frac{hv}{kT_{CMB}}$ and $\tilde{c(\nu)}\propto \nu^{-2}$.
 
 The expression for $C^f$ is then given by,
 \begin{equation}
     C^{CMB}_l(\nu_i,\nu_j)=C^{CMB}_l
 \end{equation}
 \begin{equation}
     C^{ff}_l(\nu_i,\nu_j)=\frac{\Theta^{ff}(\nu_i)\Theta^{ff}(\nu_j)}{\Theta^{ff}(31 \rm{GHz})^2}(70\rm{\mu K})^2 \ell^{-3}R(\nu_i,\nu_j,0.02)
 \end{equation}
 \begin{equation}
     C^{Dust}_l(\nu_i,\nu_j)=\frac{\Theta^{Dust}(\nu_i)\Theta^{Dust}(\nu_j)}{\Theta^{Dust}(90 \rm{GHz})^2}(24\rm{\mu K})^2 \ell^{-3}R(\nu_i,\nu_j,0.3)
 \end{equation}
 \begin{equation}
     C^{sync}_l(\nu_i,\nu_j)=\frac{\Theta^{sync}(\nu_i)\Theta^{sync}(\nu_j)}{\Theta^{sync}(19 \rm{GHz})^2}(101\rm{\mu K})^2 \ell^{-2.4}R(\nu_i,\nu_j,0.15)
 \end{equation}
 \begin{equation}
     C^{radio}_l(\nu_i,\nu_j)=\frac{\Theta^{radio}(\nu_i)\Theta^{radio}(\nu_j)}{\Theta^{radio}(150 \rm{GHz})^2}(\sqrt{3}\rm{\mu K})^2 \frac{2\pi}{\ell(\ell+1)}\left[\frac{\ell}{3000}\right]^2R(\nu_i,\nu_j,0.5)
 \end{equation}
  \begin{equation}
  \begin{split}
     C^{IR}_l(\nu_i,\nu_j)=\frac{\Theta^{IR}(\nu_i)\Theta^{IR}(\nu_j)}{\Theta^{IR}(150 \rm{GHz})^2} \frac{2\pi}{\ell(\ell+1)}(7\rm{\mu K^2}\left(\frac{\ell}{3000}\right)^2\\ +5.7\rm{\mu K^2}\left(\frac{\ell}{3000}\right)^{0.8})
     R(\nu_i,\nu_j,0.3)
     \end{split}
 \end{equation}
 
\begin{figure}
\centering 
\includegraphics[width=\columnwidth]{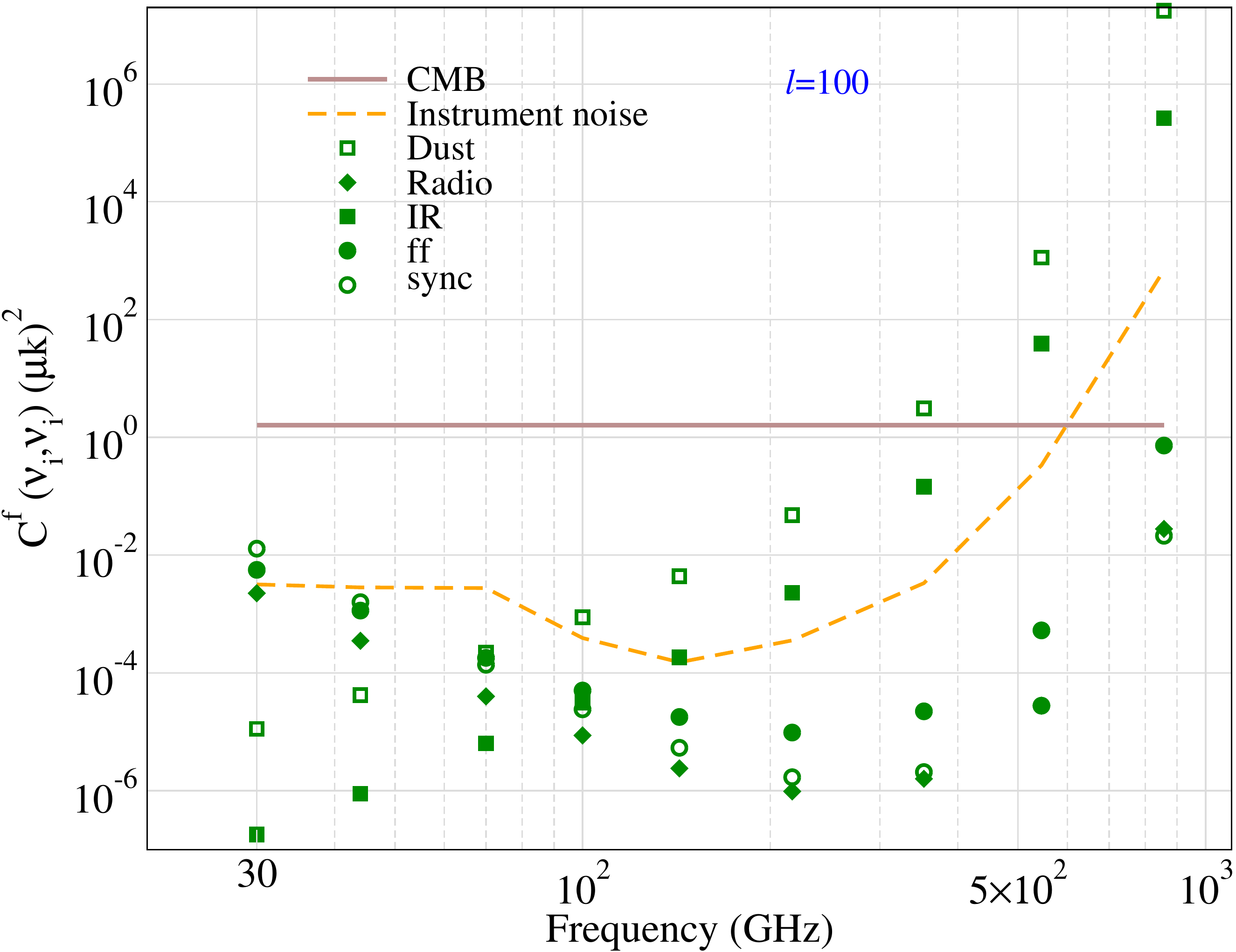}
\caption{Diagonal element of $C^{f}(\nu_i,\nu_j)$ for each foreground for tSZ extraction (ignoring ntSZ) as a function of frequency at $\ell=100$. }
\label{fig:noise_l_100}
\end{figure}
\begin{figure}
\centering 
\includegraphics[width=\columnwidth]{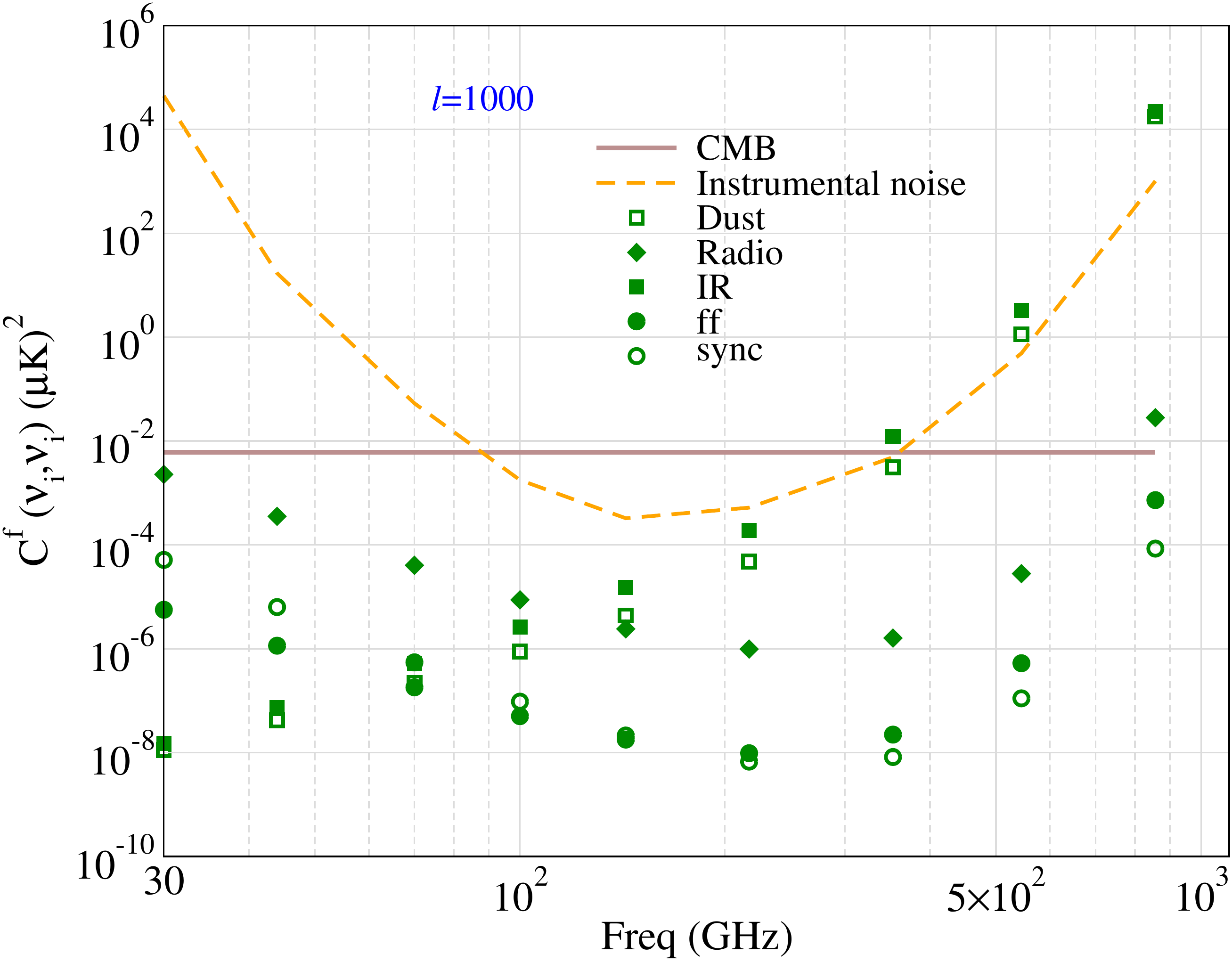}
\caption{Same as Fig \ref{fig:noise_l_100} but at $\ell=1000$.}
\label{fig:noise_l_1000}
\end{figure}

\begin{figure}
\centering 
\includegraphics[width=\columnwidth]{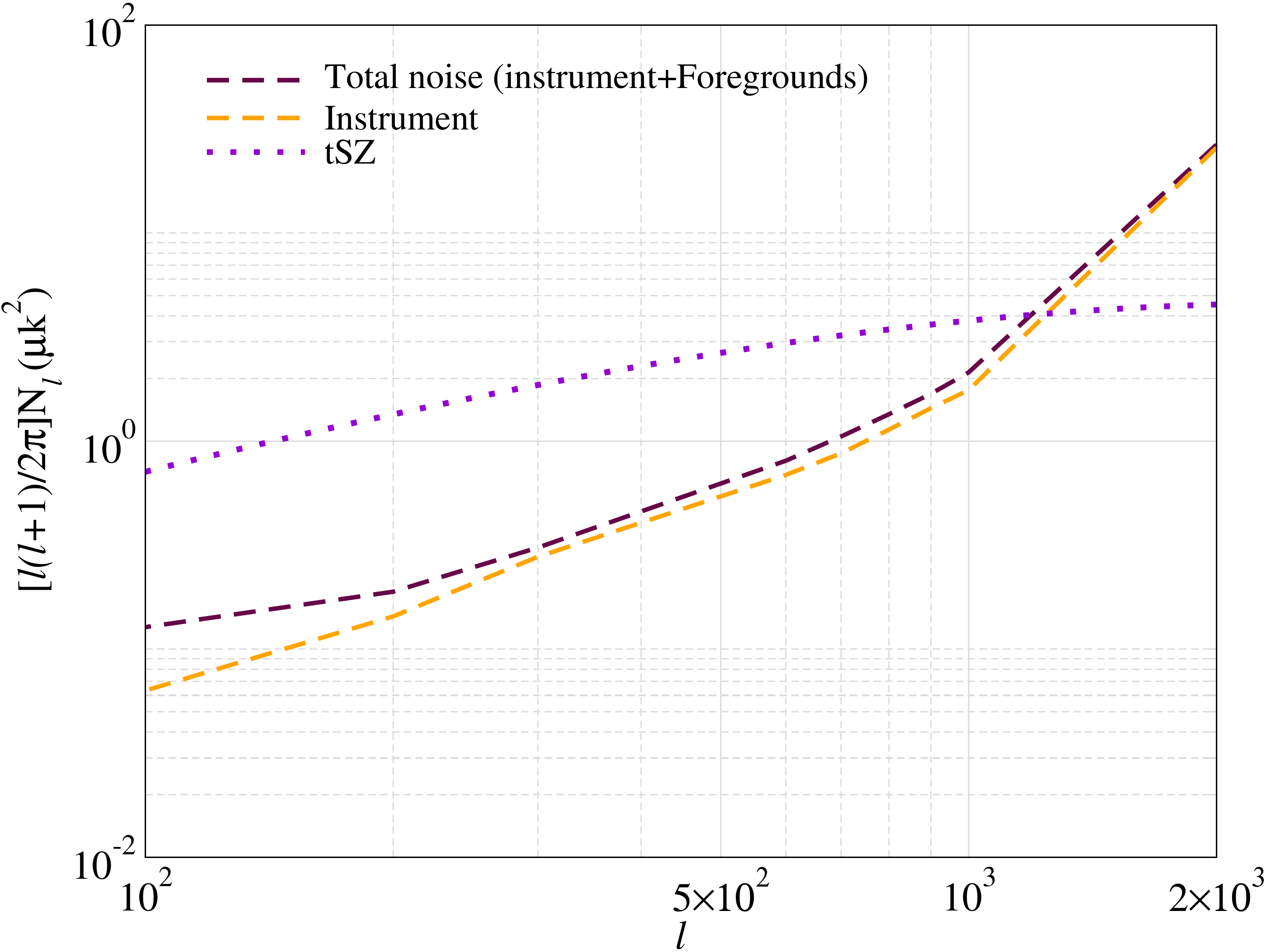}
\caption{Noise after SZ signal extraction and comparison with the tSZ signal from Planck.}
\label{fig:tsz_extraction}
\end{figure}
 
We first try to extract tSZ signal from galaxy clusters, ignoring ntSZ from radio galaxies. This exercise was done in \cite{HP2013}. We repeat the exercise as a check of our calculations. We begin by using the procedure described in Sec. \ref{sec:fluctuation_detection} to look at the prospect of detecting thermal SZ signal in Planck which has 9 frequency channels. In Fig. \ref{fig:noise_l_100} and \ref{fig:noise_l_1000}, we plot the frequency dependence of $C^f(\nu_i,\nu_i)$ for two multipole, $\ell$ = 100, 1000, using Planck instrumental noise. The dominant foregrounds are dust, IR point sources and instrumental noise at high frequencies and CMB at intermediate frequencies. At high multipoles (like $\ell = 1000$), the instrumental noise also dominates at all frequencies.

Given the frequency dependence of the foregrounds as shown in Fig. \ref{fig:noise_l_100} and \ref{fig:noise_l_1000},  we show the possibility of extracting the thermal SZ signal with Planck sensitivity and frequency bands in Fig. \ref{fig:tsz_extraction}. 
 The instrumental noise is perfectly decorrelated across frequencies and, therefore, it dominates the total noise budget. The residual CMB and the other dominant foregrounds are well below the instrumental noise as the ILC procedure is efficient in cleaning correlated noise. The thermal SZ signal is well above the noise up to $\ell \sim 1000$.  While this exercise was done with only 9 frequencies, more number of frequency channels will further reduce the residual foreground noise.

\if{
 \section{kSZ power spectrum from cosmological radio galaxy distribution}
 \color{red}(comment- I will have to recheck the calculations in this section and will add figures later. Anyway, kSZ turns out to be 6-7 magnitudes below the tSZ.\color{black})
Moving electrons boost the CMB photons through Doppler effect and give rise to kSZ effect \citep{SZ1980}. The expression for kSZ power spectrum is given by \citep{H2000,MF2002},
\begin{equation}
C_\ell=\frac{n_{e,0}^2\sigma_T^2}{H_0^2}\int dz\int \frac{dx}{x^2}(1+z)^4 e^{-2\tau}P_q(l/x,z),
\label{ksz}
\end{equation}
where $n_{e,0}$ is the background electron number density, $x$ is the comoving distance to redshift $z$, $\tau$ is the optical depth, $P_q(l/x=k,z)$ is the line of sight power spectrum. In the nonlinear regime (or high $k$ regime), we can approximate $P_q(l/x=k,z)$ as \cite{MF2002},
\begin{equation}
P_q(k,z)=\frac{2}{3}P_{\delta\delta}(k)v_{rms,lin}^2,
\label{kszpower}
\end{equation}
where $P_{\delta\delta}(k)$ is the density power spectrum and $v_{rms,lin}$ is the rms velocity in linear perturbation theory. The expression for linear rms velocity at any redshift is given by,
\begin{equation}
v_{rms,lin}(z)=H(z)^2a(z)^2f(\Omega_m)^2D(z)^2\frac{1}{2\pi^2}\int dk' P_{lin}(k'),
\end{equation}
where $D(z)$ is the growth factor and $f(\Omega_m)=\frac{dlnD}{dlna}=\Omega_m^{0.6}$ is the growth function and $P_{lin}(k')$ being the linear density power spectrum. 
The nonlinear density power spectrum $P_{\delta\delta}(k)$ is computed within the halo model. 
\par
 We assume that the ionized gas have NFW density profile. The fourier transform of the density profile is given by,
 \begin{equation}
 \rho(k,M)=\int 4\pi r^2 dr \frac{sin(kr)}{kr}\rho(r,M),
 \label{kszfourier}
 \end{equation}
 where $\rho(r)$ is the density profile of free electrons in the dark matter halo with mass $M$ and the integral is done from $r=0$ to the virial radius of the halo ($r_{vir}$). $\rho(k)$ is normalized such that $\rho(0)=1$ which is the physical condition that mass inside $r_{vir}$ is equal to the mass of dark matter halo. The two-halo term is given by \citep{S2000,CS2002},
 \begin{equation}
 P^{2h}(k)=[\int dM \frac{dn(M,z_{st})}{dM}\left(\frac{M}{\bar{\rho}}\right)b(M,z_{st})\rho(k,M)]^2 P_{lin}(k).
 \end{equation}
 The one-halo term is given by,
 \begin{equation}
  P^{1h}(k)=\int dM\frac{dn(M,z_{st})}{dM}\left(\frac{M}{\bar{\rho}}\right)^2 [\rho(k,M)]^2.
  \end{equation}
  The total non-linear density $P_{\delta\delta}(k)$ is given by,
  \begin{equation}
  P_{\delta\delta}(k)=P^{1h}(k)+P^{2h}(k).
  \end{equation}
  \par
  We use the parametric fits of \cite{SPJWFPTEC2003} for density power spectrum.
  Using the density power spectrum and rms velocity, we compute $C_\ell$ using Eq. \ref{ksz}.  For low $l$, the two halo term dominate. The two halo density power spectrum reduces to linear density power spectrum at $l<1000$ and the corresponding kSZ power spectrum was computed by \cite{OV1986}. But the one-halo term boosts the $C_\ell$ at $l\sim 1000-10000$ by a factor of few. Lower mass dark matter halos can give substantial contribution to kSZ effect as compared to tSZ and, therefore, $l(l+1)C_\ell/2\pi$ for kSZ is flat till $l\sim 10^6$ while for tSZ, $l(l+1)C_\ell/2\pi$ drops after $l\sim 3000$.

kSZ power spectrum depends upon the product of electron number density and velocity. For a radio lobe, the motion of electrons coming towards us and going away from us, cancel each other out and only the bulk motion of clusters matter. Assuming equipartition of energy inside the radio lobes, the number density of electrons inside the lobe is proportional to $p_c/m_{\rm{e}}\rm{c^2}$ \cite{KDA1997}, where $p_c$ is the pressure inside the lobe. Therefore, the expression for kSZ power spectrum for radio lobes is equivalent to tSZ power spectrum but only modulated by $v_{rms}^2$. Since, $v_{rms}\sim 10^{-3}-10^{-4}\rm{c}$ over the redshift range, kSZ power spectrum is 6-7 orders magnitude below the tSZ power spectrum.
}\fi

\end{document}